\documentclass[aps,prx,twocolumn,superscriptaddress]{revtex4-2}
\usepackage{graphicx} 
\usepackage{braket} 
\usepackage{amsmath} 
\usepackage{amssymb} 
\usepackage{bbold}
\usepackage{xcolor}
\usepackage{ragged2e}
\usepackage{microtype}
\usepackage{algpseudocode}
\usepackage{amsfonts}
\usepackage{multirow}
\usepackage{hyperref} 
\newif\ifcomments
\commentstrue

\begin{document}

\title{Hybrid quantum-classical neural network for sample-efficient recognition\texorpdfstring{\\}{ } of topological phases}

\newcommand{\affiliationethzphys}{\affiliation{Department of Physics, ETH Zurich, CH-8093 Zurich, Switzerland}}
\newcommand{\affiliationethzqc}{\affiliation{Quantum Center, ETH Zurich, CH-8093 Zurich, Switzerland}}
\newcommand{\affiliationpsiqchub}{\affiliation{ETH Zurich - PSI Quantum Computing Hub, Paul Scherrer Institute, CH-5232 Villigen, Switzerland}}
\newcommand{\affiliationerlangen}{\affiliation{Department of Physics, Friedrich-Alexander Universit{\"a}t Erlangen-N{\"u}rnberg (FAU), Erlangen, Germany}}
\newcommand{\affiliationerlangenmpi}{\affiliation{Max Planck Institute for the Science of Light, Staudtstra\ss{}e 2, 91058 Erlangen, Germany}}
\newcommand{\affiliationerlangenquint}{\affiliation{Quint Computing GmbH, Erlangen, Germany}}

\author{Markus~K.~Hoffmann}
\affiliationerlangen

\author{Leon~C.~Sander}
\affiliationerlangen

\author{Colin~Scarato}
\affiliationethzphys
\affiliationethzqc

\author{Christoph~Hellings}
\affiliationethzphys
\affiliationethzqc

\author{Johannes~Kn{\"o}rzer}
\affiliationethzphys
\affiliationethzqc
\affiliationpsiqchub

\author{Michael~J.~Hartmann}
\affiliationerlangen
\affiliationerlangenmpi
\affiliationerlangenquint

\author{Petr~Zapletal}
\affiliationerlangen

\date{\today}

\begin{abstract}
    With increasing maturity of quantum computers, standard methods for characterizing global properties of their output quantum states via direct measurements and classical post-processing are becoming increasingly impractical due to large measurement costs. Although quantum neural networks could directly process quantum states to identify underlying characteristics with reduced measurement efforts, they often require deep quantum circuits that cannot be implemented on existing devices. To overcome these challenges, we introduce a hybrid quantum-classical neural network that consists of a shallow parameterized quantum circuit, measurements, and a classical neural network. The parameterized quantum circuit performs a nonlocal transformation of the measurement basis, which is jointly trained with the classical neural network to maximize the statistical distance between data obtained by measuring different quantum states. Using supervised learning, we demonstrate that the hybrid neural network distinguishes the topological phase of the surface code from a symmetry-enriched topological phase and random product states. Moreover, this hybrid neural network reduces both inference and training sample complexities of recognizing the topological phase by approximately one order of magnitude compared to a classical neural network trained on randomized Pauli measurements. As this hybrid neural network features a shallow quantum circuit that can be readily implemented on existing quantum computers, it enables the efficient characterization of complex quantum states.
\end{abstract}

\maketitle

\section{Introduction}

Thanks to remarkable recent developments \cite{acharya2025,bluvstein2024}, quantum computers can now generate quantum states, which can no longer be fully described by classical computers \cite{morvan2024}, making the characterization of such complex quantum states a key challenge. Although many local properties of quantum states can be efficiently characterized via classical shadows based on randomized measurements \cite{huang2020,elben2023}, this approach faces an unfeasible sample complexity for global observables. An attractive alternative is to process prepared quantum states further on a quantum computer to enable a more efficient characterization \cite{biamonte2017}, including quantum principal component analysis \cite{lloyd2014}, quantum autoencoders \cite{romero2017}, certification of Hamiltonian dynamics \cite{wiebe2014,gentile021}, and quantum reservoir processing \cite{ghosh2019}. Quantum neural networks based on parameterized quantum circuits are particularly versatile in this context \cite{farhi2018,cong2019,beer2020,kottmann2021,caro2022,gong2023}.

\begin{figure*}
\centering
\includegraphics[width=0.85\linewidth]{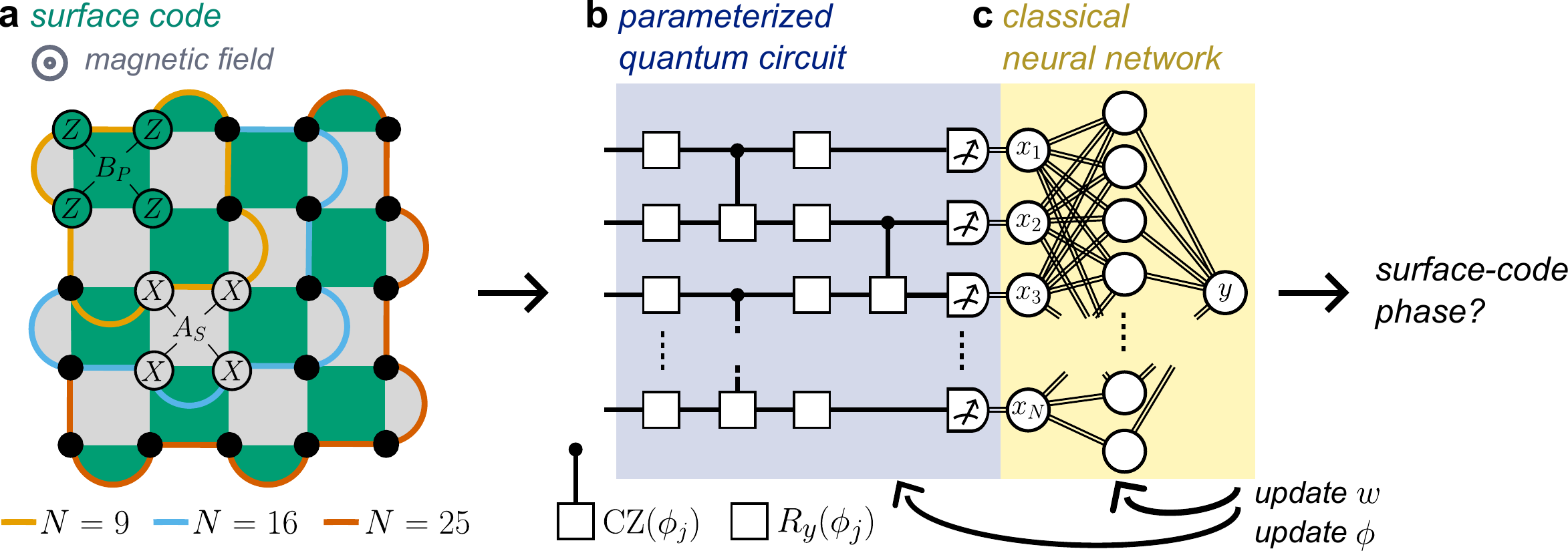}
\caption{Hybrid neural network for recognizing the topological phase of the surface code. (a) Surface code for $N=9, 16, 25$ qubits on a square lattice with qubits depicted as black points. Weight-four operators $A_s$ and $B_p$ are shown as gray and green squares, respectively. Respective weight-two operators are depicted as gray and green semicircles. To
detect the topological phase of the surface code, its ground states are processed in a hybrid quantum-classical neural network comprising a parameterized quantum circuit $U(\phi)$, measurements and a classical feedforward neural network. (b) The parameterized quantum circuit consists of single-qubit rotations $R_y(\phi_j) = \exp(-i\frac{\phi_j}{2}Y)$ with angles $\phi_j$ and Pauli operator $Y$, and two-qubit controlled phase gates $\textrm{CZ}(\phi_j)$ with phases $\phi_j$, where the measurements are performed in the computational basis. (c) The classical feedforward neural network with weights and biases $w$ assigns an output $y(x)$ to each measured bit string $x =x_1x_2\ldots x_N$.}\label{fig:hnn}
\end{figure*}

One often aims to analyze quantum many-body states generated as the output of condensed matter simulations, a promising near-term application for quantum computers \cite{satzinger2021, will2025, evered2025}, especially for two- and higher-dimensional systems that are hard to simulate using classical numerical methods \cite{zheng2017,hangleiter2020}. In particular, topological many-body systems are a subject of active research \cite{haller2023,liu2024}. In such simulations, the classification of phases of matter is a key task, which has been addressed using classical machine learning methods applied to data obtained from numerical simulations \cite{carrasquilla2017,vannieuwenburg2017}, experimental measurements \cite{rem2019,bohrdt2021,kaming2021,miles2023}, and classical shadows \cite{huang2022}. 
However, the large amounts of data required to characterize such quantum systems remain a crucial bottleneck. In particular, topological phases, which have already been realized on quantum computers \cite{satzinger2021,semeghini2021}, are especially difficult to characterize due to the absence of local order parameters.

Here, we address this open challenge by introducing a hybrid neural network, depicted in Fig.~\ref{fig:hnn}, that consists of a shallow parameterized quantum circuit, measurements and a classical feedforward neural network. Its parameterized quantum circuit implements a nonlocal transformation of the measurement basis that facilitates phase recognition by variationally maximizing the statistical distance between measurement outcomes from different phases. To jointly train the parameterized quantum circuit and the classical neural network, we develop a two-loop optimization procedure. We show that this hybrid neural network distinguishes the topological phase of the surface code, exhibiting intrinsic topological order \cite{kitaev2003}, not only from a topologically trivial phase, but also from a symmetry-enriched topological phase \cite{liu2024} and random product states forming a featureless 1-design \cite{scott2006}. Importantly, this hybrid neural network reduces both inference and training sample complexities by approximately one order of magnitude compared to a classical feedforward neural network trained on the outcomes of randomized Pauli measurements. It thus outperforms existing approaches in these aspects.

Our hybrid neural network uses a much shallower quantum circuit than quantum convolutional neural networks \cite{cong2019}. Indeed, the deep circuits of the latter have limited their investigation mainly to classical numerical simulations \cite{cong2019,caro2022,liu2023}, while proof-of-principle realizations have considered one-dimensional spin chains \cite{herrmann2022,chen2025}, which can be simulated classically \cite{mcculloch2008}. Applications of quantum convolutional neural networks to two-dimensional systems, where quantum computers promise to provide computational advantage, have only been studied theoretically \cite{sander2025,aktar2025}. In contrast, the shallow quantum circuit of our hybrid neural network enables investigations on currently available and near-term quantum computers. This is explored in the
companion paper \cite{scarato2026}.  

In some cases, quantum convolutional neural networks employ explicitly constructed, non-trainable circuits that can be significantly shortened by performing a large part of the processing on a classical computer after measuring the qubits \cite{herrmann2022,zapletal2024,sander2025}. 
Other approaches for dequantizing quantum neural networks \cite{cerezo2025,bermejo2026}, which rely on the efficient acquisition of data from quantum devices, have revealed the lack of nontrivial benchmark datasets. In many cases
\cite{cong2019,liu2023,caro2022,chen2025,herrmann2022}, quantum neural networks have been tested on ``locally easy'' datasets that can be efficiently classified by classical models using information encoded in the local order parameters of topologically trivial phases \cite{bermejo2026}.
To show that our approach works beyond these scenarios, we consider two datasets consisting solely of phases that lack local order parameters: the first comprises topological ground states of the surface code and the featureless 1-design of product states, while the second is formed by a parameterized family of states \cite{liu2024} spanning the phase transition between the surface-code phase and the symmetry-enriched topological phase.
We show that the hybrid neural network classifies these states with a reduced sample complexity compared to the classical network trained on randomized Pauli measurements.

Thanks to its clear working principles and sampling advantage, as discussed here, and the experimental realization reported in~\cite{scarato2026}, the hybrid neural network paves the way for the efficient characterization of complex quantum states arising in condensed matter physics and other applications in quantum computing.

\section{Quantum-classical hybrid neural network}
We design a hybrid neural network to recognize quantum phases. For simplicity, we consider the task of distinguishing two quantum phases with labels $L=0,1$ from one another. As depicted in Fig.~\ref{fig:hnn}, this hybrid neural network consists of a parameterized quantum circuit $U(\phi)$, measurements, and a classical feedforward neural network. To recognize the phase of a quantum state $\ket{\psi}$, we process it in the parameterized quantum circuit $U(\phi)$ and measure all qubits $i=1,2,\ldots,N$ in the computational basis. In classical post-processing, we then use a feedforward neural network to assign an output value $y(x)\in[0,1]$ to each measured bit string $x=x_1x_2\ldots x_N$, where $x_i=0,1$ corresponds to the measurement outcome of qubit $i$. The output $y(x)$ corresponds to the confidence score of phase $L=1$, whereas $1 - y(x)$ is the confidence score of phase $L=0$.

We consider the parameterized quantum circuit depicted in Fig.~\ref{fig:hnn}b, which consists of single-qubit rotations $R_y(\phi_j) = \exp(-i\frac{\phi_j}{2}Y)$ with angles $\phi_j$ and Pauli operator $Y$, and two-qubit controlled phase gates with phases $\phi_j$. All gates are parameterized independently and we use the shorthand notation $\phi = \phi_1\phi_2\ldots\phi_K$. 
These parameterized gates can be natively implemented \cite{scarato2025} on superconducting quantum processors with the architecture developed in \cite{krinner2022}, on which the hybrid neural network has been realized \cite{scarato2026}.
The classical feedforward neural network, in turn, comprises an input layer with $N$ nodes, a hidden fully-connected layer with $n$ nodes and rectified linear unit activation functions, and an output layer with a single node using a sigmoid activation function. We use a shorthand notation $w$ for the weights and biases of this feedforward network. 

In this work, we focus on supervised learning, which has previously been studied using both classical neural networks \cite{carrasquilla2017} and quantum neural networks \cite{cong2019,caro2022}. For each phase $L=0,1$, the training data comprise $M_L$ quantum states $\ket{\psi_L^{(m)}}$, where $m=1,2,\ldots,M_L$ and $M_0\neq M_1$ in general. The hybrid neural network is trained by minimizing the binary cross-entropy cost function
\begin{align}\label{eq:bxe}
    C =-\sum_{L=0,1}\frac{1}{2M_{L}} \sum_{m=1}^{M_L}&\frac{1}{|\mathcal{D}_L^{(m)}|}\sum_{x\in \mathcal{D}_L^{(m)}}\{L\ln[y(x)]\nonumber\\
    &+ (1 - L)\ln[1 - y(x)]\},
\end{align}
with respect to $\phi$ and $w$, where each dataset of bit strings $\mathcal{D}_L^{(m)}$ is obtained by the repeated processing and measurement of the state $\ket{\psi_L^{(m)}}$.

{
\renewcommand{\thetable}{\arabic{table}}
\renewcommand{\tableautorefname}{Algorithm}
\begin{table}[t]
\refstepcounter{table}
\label{tab:train}
\begin{ruledtabular}
\begin{tabular}{p{0.98\columnwidth}}
\textbf{Algorithm 1: Hybrid neural network training}\\
\hline
\begin{algorithmic}[1]
\State \textbf{Input:} Training states $\ket{\psi_L^{(m)}}$ for each phase $L=0,1$
\State \textbf{Output:} Optimized parameters $\phi^*$ of the quantum circuit, and weights and biases $w^*$ of the feedforward neural network
\State Randomly initialize $\phi^{(1)}$ and $w^{(1)}$
\State \textbf{Outer Loop:}
\For{$k = 1,2,\ldots, N_{\rm it}$}
\State  \parbox[t]{0.86\columnwidth}{\justifying \noindent 
Process training states $\ket{\psi_L^{(m)}}$ by the parameterized circuit $U(\phi^{(k)})$ and measure sets of bit strings $\mathcal{D}_L^{(m)}$ }
\State \textbf{Inner loop:} 
\State \indent \parbox[t]{0.78\columnwidth}{\justifying \noindent
Train feedforward neural network $w^{(k)}$ on sets $\mathcal{D}_L^{(m)}$ and return residual cost $C^{(k)}_{\rm res} = \min_{w} C^{(k)}$
}
\State \indent Update weights and biases $w^{(k+1)} \gets w^{(k)}$
\State \parbox[t]{0.86\columnwidth}{\justifying \noindent
Estimate gradient $\textrm{grad}_{\phi} C^{(k)}_{\rm res}$ of the residual cost}
\State \parbox[t]{0.86\columnwidth}{\justifying \noindent
{Update parameters $\phi^{(k+1)} \gets \phi^{(k)} -\alpha\, \textrm{grad}_{\phi} C^{(k)}_{\rm res}$} with learning rate $\alpha$}
\EndFor
\State Set $\phi^* \gets\phi^{(j)}$ and $w^* \gets w^{(j)}$, where $j=\arg\min_{k=1,2,\ldots,N_{\rm it}}C^{(k)}_{\rm res}$
\end{algorithmic}
\end{tabular}
\end{ruledtabular}
\end{table}
}
\setcounter{table}{0}

To jointly train the classical neural network and the parameterized quantum circuit, we introduce Algorithm~\ref{tab:train}, which comprises two optimization loops. In the inner loop, we train the classical neural network on datasets of bit strings $\mathcal{D}_L^{(m)}$ measured for given parameters $\phi$ of the quantum circuit. In the outer loop, we minimize the residual cost $C_{\rm res} = \min_{w} C$ with respect to the parameters $\phi$ of the quantum circuit.
To numerically simulate the processing of quantum states $\ket{\psi_L^{(m)}}$ in the circuit $U(\phi)$, we use the QuTiP Python package \cite{johansson2013}.
For optimizing the parameters $\phi$ of the quantum circuit in the outer loop, we choose the evolution strategy optimizer \cite{salimans2017}, which estimates the gradient of the residual cost from stochastic perturbations $\delta\phi$ of parameters $\phi$. Finally, for training the feedforward neural network in the inner loop of Algorithm~\ref{tab:train}, we use the Keras API \cite{chollet2015} integrated in TensorFlow. 

The mapping of measured bit strings $x$ to the output value $y(x)$ is nonlinear due to the parameterization by the feedforward neural network. However, the mean output
\begin{equation}\label{eq:mean}
\bar{y} = \frac{1}{|\mathcal{D}^{(m)}_L|} \sum_{x\in\mathcal{D}^{(m)}_L} y(x)    
\end{equation}
corresponds to an expectation value $\lim_{|\mathcal{D}^{(m)}_L|\rightarrow\infty}\bar{y}=\textrm{Tr}[O\ket{\psi_L^{(m)}}\bra{\psi_L^{(m)}}]$ of the quantum observable $O=\sum_{x\in\{0,1\}^N}y(x)U^{\dagger}(\phi)\ket{x}\bra{x}U(\phi)$, which is a linear function of the input quantum state $\ket{\psi_L^{(m)}}$, where $\ket{x}$ is the computational basis eigenstate corresponding to $x$. Similarly, the cost function $C$ corresponds to an expectation value of a quantum observable averaged over all $M_0 + M_1$ training states.

\section{Working principles of the hybrid neural network}\label{sec:theory}

Before learning specific quantum phases, we discuss the roles played by the classical feedforward neural network and the parameterized quantum circuit in our quantum phase recognition approach. We start with the analytical analysis of the classical neural network \cite{arnold2022,arnold2023}, assuming fixed parameters of the quantum circuit for now. 

The feedforward neural network estimates the probability 
\begin{align}\label{eq:yopt}
    y(x)\approx y_{\rm opt}(x) = \frac{P_1(x)}{P_0(x) + P_1(x)},
\end{align}
that the observed bit string $x$ originates from phase $L=1$, assuming that both phases are sampled uniformly, where
\begin{equation} \label{eq:probclass}
P_L(x) = \frac{1}{M_L}\sum_{m=1}^{M_L}P_L^{(m)}(x),
\end{equation}
and $P_L^{(m)}(x)=|\bra{x}U(\phi)\ket{\psi_L^{(m)}}|^2$ is the probability to measure bit string $x$ after having processed the state $\ket{\psi_L^{(m)}}$ in the quantum circuit $U(\phi)$. 

The probability $y_{\rm opt}(x)$ is estimated in a data-driven manner by minimizing the cost function \eqref{eq:bxe} with respect to $w$, where $y_{\rm opt}(x)$ is the global minimum. Thanks to the universal function approximation capability of the neural network, its output $y(x)$ approaches this global minimum in the limit of a large number of measurement shots $|\mathcal{D}_L^{(m)}|\rightarrow\infty$, for all $m$ and $L$, provided that the network is sufficiently expressive and well trained \cite{arnold2022}. The residual cost
\begin{align}\label{eq:bxe_opt}
    C_{\rm res}\approx C_{\rm opt} = \ln{2} - D_{JS}[P_0,P_1]
\end{align}
arises from the overlap between the probability distributions $P_0$ and $P_1$, where the Jensen-Shannon divergence $D_{JS}[P_0,P_1] = \frac{1}{2}\sum_{L=0,1}\sum_{x\in\{0,1\}^N}P_L(x)\ln\frac{P_L(x)}{Q(x)}$ measures their statistical distance and $Q(x) =  \frac{1}{2}[P_0(x) + P_1(x)]$.

For such a well-trained neural network, the error rate 
\begin{equation}
    p_{\rm err} = \sum_{L=0,1}\frac{1}{2M_{L}} \sum_{m=1}^{M_L}\frac{1}{|\mathcal{D}_L^{(m)}|}\sum_{x\in \mathcal{D}_L^{(m)}}\textrm{err}[y(x),L]
\end{equation}
is close to the \textit{Bayes error} rate
\begin{equation}
    p_{\rm opt} = \frac{1}{2}\sum_{\substack{x\in\{0,1\}^N \\ P_0(x)\leq P_1(x)} }P_0(x) + \frac{1}{2}\sum_{\substack{x\in\{0,1\}^N \\ P_0(x)> P_1(x)} }P_1(x),
\end{equation}
where errors are defined as $\textrm{err}[y,L] = |L - \theta[y-\frac{1}{2}]|$ and $\theta$ is the Heaviside step function. The Bayes error rate $p_{\rm opt} = \frac{1}{2} \left(1 - \textrm{TV}[P_0,P_1]\right)$ also stems from the overlap between the probability distributions $P_0$ and $P_1$ \cite{arnold2023}, where $\textrm{TV}[P_0,P_1] = \frac{1}{2}\sum_{x\in\{0,1\}^N}|P_0(x) - P_1(x)|$ denotes the total variation distance.

\begin{figure}
\centering
\includegraphics[width=\linewidth]{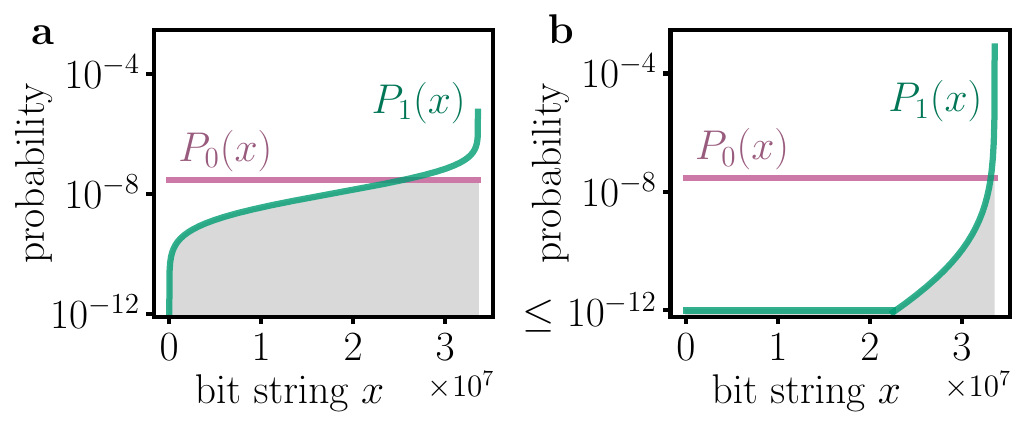}
\caption{Probability distribution $P_1(x)$ of bit strings measured for the topological ground states of the surface code processed in the quantum circuit (green line) for (a) parameters $\phi^{(1)}$ initialized uniformly at random and (b) trained parameters $\phi^*$, compared to the featureless uniform distribution $P_0(x)=2^{-N}$ of a 1-design (purple line). The parameterized quantum circuit is trained to minimize the overlap (gray area) between these probability distributions. Here, we order the bit strings $x$ separately in each panel according to the values of $P_1(x)$, from smallest to largest.}\label{fig:overlap}
\end{figure}

As will become evident from the concrete examples below, the overlap between the underlying probability distributions of the training data can severely limit the accuracy of recognizing quantum phases. Large overlaps of probability distributions typically occur for topological phases whose characteristic nonlocal quantum correlations are hard to estimate from measurement outcomes $x$. These are observed for the quantum circuit $U(\phi^{(1)})$ with parameters $\phi^{(1)}$ randomly chosen from the interval $[0,2\pi)$ as shown in Fig.~\ref{fig:overlap}a and experimentally demonstrated in the companion paper \cite{scarato2026}. Large overlaps are also observed for state-of-the-art randomized measurement approaches \cite{huang2022,bermejo2026}, as discussed in Sec.~\ref{sec:complexity}.

Here, we address this open problem by optimizing the parameterized quantum circuit $U(\phi)$ to increase the statistical distance $D_{JS}[P_0,P_1]$ of the data generated by repeated processing and measurement of the training quantum states $\ket{\psi_L^{(m)}}$, as shown in Fig.~\ref{fig:overlap}b. Using relation \eqref{eq:bxe_opt} and the well-trained feedforward neural network, we achieve this in a data-driven manner by minimizing the residual cost $C_{\rm res}$ with respect to the parameters $\phi$ of the quantum circuit $U(\phi)$.

To summarize, the classical feedforward neural network estimates the probability $y_{\rm opt}(x)$ and achieves an error rate close to the Bayes error rate, $p_{\rm err} \approx p_{\rm opt}$, which is only limited by the statistical overlap of the measured data. In turn, the parameterized quantum circuit is trained to increase the statistical distance between the resulting measurement distributions, thereby reducing the Bayes error rate. This is achieved by the joint training of the feedforward neural network and the parameterized quantum circuit, see Algorithm~\ref{tab:train}.

\section{Surface code in magnetic field}
Having introduced the hybrid neural network, we now use it to recognize quantum phases of the toric code \cite{kitaev2003}. In particular, we consider its realization on a plane, referred to as the surface code \cite{satzinger2021}, with qubits located on a two-dimensional square lattice as shown in Fig.~\ref{fig:hnn}a. It is described by the Hamiltonian
\begin{equation}\label{eq:sc}
    H_{\rm SC} = -\sum_s A_s -\sum_p B_p,
\end{equation}
where $A_s = \prod_{i\in s}X_i$ and $B_p = \prod_{i\in p}Z_i$ are products of four Pauli operators shown in Fig.~\ref{fig:hnn}a as gray and green squares, respectively, and $X_i$ and $Z_i$ are Pauli operators on qubit $i$. Moreover, we consider open boundary conditions involving weight-two operators $A_s$ and $B_p$ shown as gray and green semicircles, respectively. Note that such weight-two boundary operators are also common in quantum error correction codes \cite{acharya2025, bluvstein2024}.

The surface code is a paradigmatic example of a system featuring $\mathbb{Z}_2$ topological order. When perturbed by a magnetic field of strength $h$,
\begin{equation}\label{eq:field}
    H = H_{\rm SC} -h\sum_{i=1}^N Z_i,
\end{equation}
it undergoes a second-order phase transition from the topological phase to a topologically trivial magnetic phase \cite{trebst2007}. For system sizes that can be solved with classical numerical methods, the phase boundary can be obtained as a dip in the second-order derivative $\textrm{d}^2 E/\textrm{d}h^2$ of the ground state energy $E$ with respect to $h$. For $N=25$, we find $h_c=0.3$ via exact diagonalization \footnote{The ground state is two-fold degenerate for $h=0$ and the degeneracy is lifted by a finite magnetic field $h>0$. For $h=0$, we consider the ground state continuously connected to the ground state family for $h>0$ in the limit $h\rightarrow 0^+$.}. 
The ground states of $H$ exhibit large magnetization, $M = \frac{1}{2N}\sum_{i=1}^N \langle Z_i \rangle$, in the trivial phase for $h>h_c$ and small magnetization in the topological phase for $h<h_c$ \cite{trebst2007}. Since magnetization can be efficiently learned, for example, via training on the outcomes of direct randomized measurements \cite{huang2022}, we expect that these ground states can be classified by classical machine learning models, similarly to locally easy datasets in one-dimensional systems \cite{bermejo2026}.

To show the capabilities of the hybrid neural network, we instead train it to distinguish the topological ground states ($h<h_c$) from a 1-design \cite{scott2006}, i.e., an ensemble of states $\ket{\psi_0^{(m)}}$ satisfying $\frac{1}{M_0}\sum_{m=1}^{M_0}\ket{\psi_0^{(m)}}\bra{\psi_0^{(m)}}=\frac{1}{2^{N}}I$, where $I$ is the identity operator. To classify these states, the hybrid neural network has to uncover topological features of the surface-code phase as the 1-design $\ket{\psi_0^{(m)}}$ exhibits a featureless uniform distribution of measurement outcomes, $P_0(x) = 2^{-N}$ for all $x$, after applying any parameterized circuit $U(\phi)$. The ensemble of product states $\mathcal{P}_N = \{|0\rangle,|1\rangle,|+\rangle,|-\rangle,|+i\rangle,|-i\rangle \}^{\otimes N}$, which are eigenstates of Pauli strings $\{X,Y,Z\}^{\otimes N}$, is a prominent example of a 1-design \footnote{The six eigenstates $\mathcal{P} = \{|0\rangle,|1\rangle,|+\rangle,|-\rangle,|+i\rangle,|-i\rangle \}$ of the Pauli operators form a 1-design of a single qubit as $\frac{1}{6}\sum_{\ket{\psi}\in \mathcal{P}} \ket{\psi}\bra{\psi} = \frac{1}{2}I$. Their tensor products $\mathcal{P}_N = \{|0\rangle,|1\rangle,|+\rangle,|-\rangle,|+i\rangle,|-i\rangle \}^{\otimes N}$ yield an $N$-qubit 1-design, since $\frac{1}{6^N}\sum_{\ket{\psi}\in \mathcal{P}_N} \ket{\psi}\bra{\psi} = \bigotimes_{j=1}^N\left(\frac{1}{6}\sum_{\ket{\psi_{j}}\in \mathcal{P}} \ket{\psi_{j}}\bra{\psi_{j}} \right)=\frac{1}{2^N}I$.}. Crucially, these states do not need to be processed in the parameterized quantum circuit $U(\phi)$. Instead, we directly draw the measurement outcomes $x$ from the uniform distribution $P_0(x)$ using a pseudorandom number generator.

 In contrast to local order parameters such as magnetization, learning features of the surface-code phase, including the topological entanglement entropy \cite{kitaev2006} and characteristic correlations \cite{cong2022}, requires a large amount of measurement data, since these features are nonlocal. For example, when using classical shadows based on randomized measurements \cite{huang2020}, the sample complexity of learning the surface-code phase grows exponentially with the length of correlations in the system \cite{huang2022}. The correlation length is typically large for physically relevant parameter regimes, for example, close to critical points of quantum phase transitions. 

Detecting characteristic features of topological order, such as Wilson loops for the surface code, can be facilitated by processing its ground states, c.f. Eq.~\eqref{eq:field}, through a quantum circuit.
For example, the non-trainable convolutional layer of the quantum neural network, which was explicitly constructed in Ref.~\cite{sander2025}, performs a transformation into the simultaneous eigenbasis of all $A_s$ and $B_p$ operators, enabling the efficient measurement of Wilson loops at all length scales and hence the recognition of the topological phase from only a few measurement shots. This convolutional layer was constructed by inverting the preparation circuit \cite{satzinger2021} for the ground state of the unperturbed surface code, c.f. Eq.~\eqref{eq:sc}. 

\section{Topological phase recognition}\label{sec:tpr}

We now proceed to learning such a nonlocal basis transformation by training the parameterized quantum circuit $U(\phi)$ using Algorithm~\ref{tab:train}. We design the parameterized circuit such that, for a specific set of parameters $\phi_{\rm inv}$ given in Appendix~\ref{sec:circ}, it implements the inverse of the preparation circuit for the unperturbed surface code ($h=0$) to allow for direct comparison with the measurement in the simultaneous eigenbasis of all $A_s$ and $B_p$ operators. 

The training data for the surface-code phase, $L=1$, comprise three topological ground states $\ket{\psi_1^{(m)}}$ ($m=1,2,3$) for $N=25$ qubits and field strengths $h=0,0.1,0.2$. The quantum circuit features $K=109$ parameters. We train the feedforward neural network with $n=128$ hidden nodes for 1000 epochs in the first iteration of Algorithm~\ref{tab:train} and for 10 epochs in all subsequent iterations on $|\mathcal{D}_1^{(m)}|=2^{16}$ bit strings measured for each state $\ket{\psi_1^{(m)}}$ and $3\times2^{16}$ bit strings drawn from the 1-design. See Appendix \ref{sec:train} for details of the hybrid neural network training and the tuning of the hyperparameters of the feedforward neural network and the evolution strategy optimizer.
The evolution strategy optimizer in the outer loop of Algorithm~\ref{tab:train} typically converges within 200 iterations to the residual cost $C_{\rm res} = 0.04 \pm 0.01$ exhibiting variations across 20 random initializations of the hybrid neural network, due to local minima in the cost landscape.

As explained in Sec.~\ref{sec:theory}, we maximize the statistical distance of data $x$ obtained by measuring the topological ground states $\ket{\psi_1^{(m)}}$ from the uniformly distributed data drawn from the 1-design by minimizing the residual cost $C_{\rm res}$ with respect to the quantum circuit parameters $\phi$. The corresponding probability distributions $P_1(x)$ and $P_0(x)$ are plotted in Figs.~\ref{fig:overlap}a and ~\ref{fig:overlap}b for a particular instance of randomly initialized parameters $\phi^{(1)}$ and the corresponding trained parameters $\phi^*$. We can see that the overlap between the two distributions (gray area) is substantially reduced in training. As a result, the error rate is reduced to $p_{\rm err}= (1.6 \pm 0.4)\%$ from $p_{\rm err}= (23 \pm 3)\%$, where the latter is achieved by training only the feedforward neural network for the initial random parameters of the quantum circuit $\phi^{(1)}$.

When we restrict the training data for the surface-code phase to the ground state for $h=0$, the inverse preparation circuit $U(\phi_{\rm inv})$ reaches the global minimum of the residual cost $C_{\rm opt} = \frac{\ln(2)}{2}\, N 2^{-N} + \mathcal{O}(2^{-N})$ and the error rate $ p_{\rm opt}=1/2^{N+1}$, which rapidly decrease with increasing system size $N$ and attain near-zero values $C_{\rm opt} =3\times10^{-7}$ and $ p_{\rm opt}=10^{-8}$ for $N=25$.  In turn, for the entire surface-code training set comprising the three ground states at field strengths $h=0,0.1,0.2$, the inverse preparation circuit $U(\phi_{\rm inv})$ reaches the residual cost $C_{\rm opt} = 9\times10^{-5}$. While this is smaller than the minimal cost $C_{\rm res} = 0.02$ achieved in training, both the inverse preparation circuit and the trained circuit facilitate the recognition of the surface-code phase with reduced error rates $p_{\rm opt} = 0.002\%$ and $p_{\rm err} = 0.7\%$, respectively. 

\begin{figure}
    \centering
    \includegraphics[width=\linewidth]{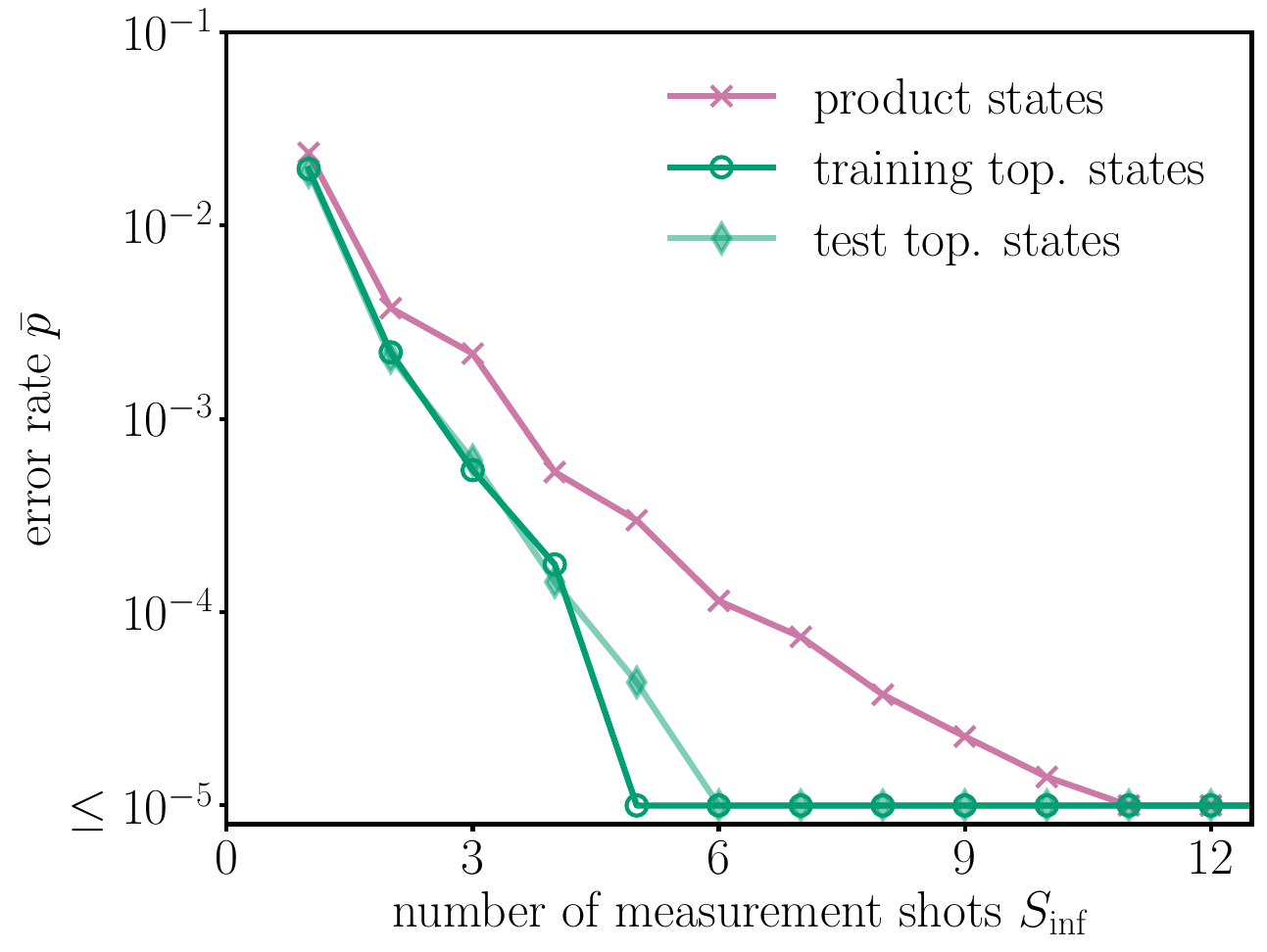}
    \caption{Multi-shot inference for recognizing the topological phase of the surface code. The false negative error rate $\bar{p}_{\rm fn}$ for the training topological states (circles) and test topological states (diamonds), as well as the false positive error rate $\bar{p}_{\rm fp}$ for $10^6$ random product states (crosses) as a function of the measurement shot number $S_{\rm inf}$.  
    }
    \label{fig:inference}
\end{figure}

We now study the inference performance of the trained hybrid neural network on various test quantum states.
We start with test topological states that are the ground states of the surface code for field strengths $h=0.05,0.15,0.25$, which are unseen during training. The rate $p_{\rm fn} = (1.6\pm0.4)\%$ of false negative errors $\textrm{err}[y(x),1]$ for single measurement shots $x$ is close to the false negative error rate $p_{\rm fn} = (1.5 \pm 0.4)\%$ for the training states for $h=0,0.1,0.2$, showing that the trained hybrid neural network generalizes well to these topological test states. To further reduce the error rate for the hybrid neural network trained for a specific random initialization, we perform a multi-shot inference by calculating the mean, see  Eq.~\eqref{eq:mean}, using multiple measurement shots for each topological state $\ket{\psi^{(m)}_1}$. We plot in Fig.~\ref{fig:inference} the rate $\bar{p}_{\rm fn}$ of multi-shot false negative errors $\textrm{err}[\bar{y},1]$ for both training topological states (circles) and test topological states (diamonds) as a function of the shot number $S_{\rm inf}=|\mathcal{D}^{(m)}_1|$ in inference. As few as $S_{\rm inf}=5$ shots are sufficient to achieve near-perfect classification, with $\bar{p}_{\rm fn}<0.01\%$. 

Next, we classify $10^6$ product states that are randomly chosen from the ensemble $\mathcal{P}_N$. All these product states are topologically trivial and feature vanishing expectation values for most products of the operators $A_s$ and $B_p$. Therefore, they enable us to test whether the hybrid neural network learned genuine features of the surface-code phase. The rate of single-shot false positive errors $\textrm{err}[y(x),0]$ for these product states is $p_{\rm fp} = 2.4 \%$ evaluated for a specific random initialization of the hybrid neural network. Additionally, we plot the rate of multi-shot false positive errors $\textrm{err}[\bar{y},0]$ in Fig.~\ref{fig:inference} (crosses). We achieve $\bar{p}_{\rm fp}<0.01\%$ with only moderately larger shot number $S_{\rm inf}=7$ compared to near-perfect classification of the topological states. For $S_{\rm inf}=10^4$ shots, only two of the random product states are misclassified.

In summary, the trained hybrid neural network correctly classifies the topological ground states and $10^6$ random product states, showing that it learned genuine topological features. It achieves error rates $\bar{p}_{\rm fn}<0.01\%$ and $\bar{p}_{\rm fp}<0.01\%$ in inference with only a few measurement shots. Moreover, the network can also classify topologically trivial ground states for $h>h_c$ after retraining only the classical feedforward neural network on the measurement outcomes $x$ taken from these states, see Appendix~\ref{sec:trivial}.

\section{Symmetry-enriched topological phase}\label{sec:set}
We now use the hybrid neural network to distinguish two topological phases, which both lack local order parameters and thus require the detection of genuine topological features. 
In particular, we focus on the surface-code phase and a symmetry-enriched topological phase by considering the family of tensor-network states
\begin{align}
    \ket{\psi} = \sum_{x \in \{0,1\}^N} \textrm{tTr}\left[T^{x_1,x_2}(g), \ldots,  T^{x_{N-1},x_N}(g) \right]\ket{x}, 
    \label{eq:tn-states}
\end{align}
where $\ket{x} = \ket{x_1x_2\ldots x_N}$ and $\textrm{tTr}$ denotes the tensor contraction. We consider rank-6 tensors $T^{x_i,x_{i+1}}(g)$, parameterized by a common parameter $ g \in [-1,1]$, introduced in Ref.~\cite{liu2024}. These states respect the anti-unitary $\mathbb{Z}_2^T$ symmetry composed of the global spin flip $\prod_{i=1}^N X_i$ and complex conjugation. The state for $g=1$ coincides with the ground state of the surface-code Hamiltonian \eqref{eq:sc}, albeit with different open boundary conditions than in Fig.~\ref{fig:hnn}a, see Appendix~\ref{sec:set_app} for details. For $g<0$, the states (\ref{eq:tn-states}) are ground states of a Hamiltonian exhibiting the symmetry-enriched topological phase. Similarly to the surface code, they feature intrinsic $\mathbb{Z}_2$ topological order but are additionally enriched by the $\mathbb{Z}_2^T$ symmetry. The surface-code phase for $g>0$ and the symmetry-enriched topological phase for $g<0$ can be distinguished by a membrane order parameter, which is a global observable consisting of the $X_i$ operator on all qubits in the bulk of the system \cite{liu2024}.

To numerically simulate these tensor-network states, we use the exact preparation circuits for $N=24$ qubits, as introduced in Ref.~\cite{liu2024}. The training data for the surface-code phase, $L=1$, and the symmetry-enriched topological phase, $L=0$, then comprise five states $\ket{\psi_L^{(m)}}$ ($m=1,2,3,4,5$) with $g$ values drawn from uniform distributions on the intervals $(0,1]$ and $[-1,0)$, respectively. As in the previous section, we construct the quantum circuit of the hybrid neural network such that it implements the inverse of the preparation circuit for the surface code ($g=1$), as a special case for the parameters $\phi_{\rm inv}$ stated in Appendix~\ref{sec:circ}.

Here, we train the feedforward neural network with $n=128$ hidden nodes for 1000 epochs in the first iteration of Algorithm~\ref{tab:train} and 30 epochs in all subsequent iterations on $|\mathcal{D}_L^{(m)}|=2^{16}$ bit strings measured for each state, see Appendix \ref{sec:train} for details.
In training, we reach the residual cost $C_{\rm res} = 0.23 \pm 0.01$ and error rate $p_{\rm err}= (8.9 \pm 0.6)\%$ whereas the inverse preparation circuit $U(\phi_{\rm inv})$ achieves a smaller cost $C_{\rm opt} = 0.12$ and error rate $p_{\rm opt} = 4.8\%$. 

\begin{figure}
    \centering
    \includegraphics[width=\linewidth]{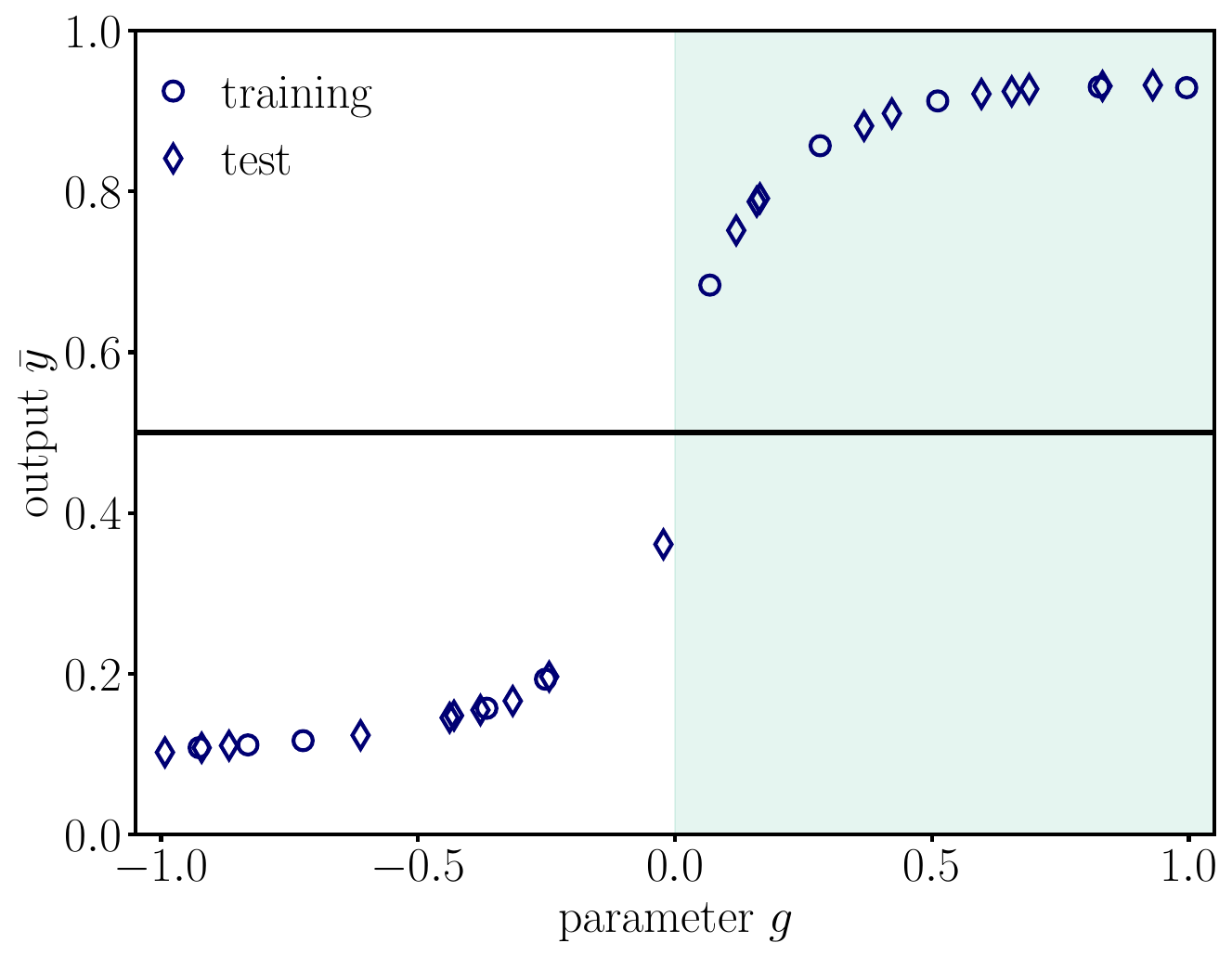}
    \caption{Distinguishing the surface-code phase from the symmetry-enriched topological phase of the family of tensor-network states introduced in Ref.~\cite{liu2024}. Mean output $\bar{y}$ as a function of the parameter $g$ for tensor-network states from the training set (circles) and test set (diamonds). The shaded region corresponds to the surface-code phase and the decision threshold $t=0.5$ is shown as the black line.}
    \label{fig:tns_output}
\end{figure}

Fig.~\ref{fig:tns_output} shows the mean output $\bar{y}$ of the trained hybrid neural network as a function of the parameter $g$. One can see that the hybrid neural network correctly classifies all states, including states for $g$ values unseen during training (diamonds). This shows that by detecting genuine features of the surface-code phase, the hybrid neural network can distinguish it from the symmetry-enriched topological phase. Moreover, the trained hybrid neural network achieves near-perfect classification, with $\bar{p}_{\rm fp},\, \bar{p}_{\rm fn}<0.01\%$ with only 13 shots, see Appendix~\ref{sec:set_app}.

\section{Sample complexity of the topological phases}\label{sec:complexity}

Finally, we benchmark the hybrid neural network against a classical feedforward neural network trained on direct randomized measurements of the topological states, i.e., without performing any quantum circuit. First, we compare the inference sample complexities of the trained networks, i.e., the number of measurement shots $S_{\rm inf }$ required to classify the topological states and product states. Then, we investigate the training sample complexities for both approaches.

To benchmark the hybrid neural network, we train the classical neural network, with five hidden fully-connected layers each comprising $n=256$ nodes, on the outcomes of randomized measurements, similarly to Ref.~\cite{huang2022}. In particular, we consider readout in local Pauli bases, directly implementable on most quantum devices via local single-qubit basis changes, which we regard as part of the measurement procedure rather than as a quantum circuit.
These randomized Pauli measurements form a positive operator-valued measure (POVM) known as the Pauli-6 POVM \cite{carrasquilla2019}. Even though these randomized measurements can be used to build classical shadows and estimate nonlinear functions of the measured quantum states \cite{huang2020}, the classical benchmark network is directly trained on the outcomes of the Pauli-6 POVM measurements. Therefore, it learns expectation values of observables, which are linear functions of the measured states, see Appendix~\ref{sec:cnn} for details. As a result, it represents an appropriate benchmark for the hybrid neural network, which also learns quantum observables.

When trained on randomized measurements of the \emph{surface code in a magnetic field} and the 1-design of product states, the classical neural network only achieves a single-shot error rate of $p_{\rm err}= (23.3 \pm 0.5)\%$, which is significantly larger than the error rate $p_{\rm err}= (1.6 \pm 0.4)\%$ of the hybrid neural network discussed in Sec.~\ref{sec:tpr}. Furthermore, $S_{\rm inf}=50$ shots are required to reach a false positive error rate of $\bar{p}_{\rm fp}<0.01\%$ in inference for this purely classical network, whereas only $S_{\rm inf}=7$ shots suffice for the hybrid neural network. The larger single-shot error rate and inference sample complexity of the classical benchmark network are caused by a substantially smaller statistical distance $D_{JS}[P_0,P_1] = 0.34$ of the Pauli-6 POVM distributions as compared to the nearly-maximal distance $D_{JS}[P_0,P_1] \approx \ln 2 - C_{\rm res} = 0.65\pm0.01$ of the measurements after processing the quantum states in the trained quantum circuit $U(\phi^*)$, see Appendix~\ref{sec:cnn} for details.

Similarly, the classical benchmark network only achieves a single-shot error rate of $p_{\rm err}= (28.43 \pm 0.07)\%$ when trained on randomized measurements of the tensor-network states spanning the \emph{transition between the surface-code and symmetry-enriched topological phases}. This is also significantly larger than the error rate $p_{\rm err}= (8.9 \pm 0.6)\%$ of the hybrid neural network discussed in the previous section. Furthermore, $S_{\rm inf}=52$ shots in inference are needed for near-perfect classification with the multi-shot error rates $\bar{p}_{\rm fp},\,\bar{p}_{\rm fn}<0.01\%$ for this classical network, whereas only $S_{\rm inf}=13$ shots are sufficient for the hybrid neural network.

To evaluate the performance of the hybrid neural network further, we study its training sample complexity. The total number of measurement shots used in training is 
\begin{equation}\label{eq:budget}
S_{\rm tot} = \sum_{L=0,1} \sum_{m=1}^{M_L}\left(N_{\rm pop}\,N_{\rm it}|\mathcal{D}_L^{(m)}| + |\mathcal{D}_{L,\textrm{re}}^{(m)}|\right),
\end{equation}
where $N_{\rm pop}$ is the number of cost function evaluations in each of the $N_{\rm it}$ iterations of the evolution strategy optimizer in the outer loop of Algorithm~\ref{tab:train} and the additional datasets $\mathcal{D}_{L,\textrm{re}}^{(m)}$ are used to retrain the classical neural network parameters $w$ after having trained the parameterized quantum circuit $U(\phi)$. The dominant contribution to the total measurement budget, $\sum_{L=0,1} \sum_{m=1}^{M_L}N_{\rm pop}\,N_{\rm it}|\mathcal{D}_L^{(m)}|$, arises from training the parameterized quantum circuit, which requires repeatedly processing the quantum states in the circuit and performing measurements, as indicated in Algorithm~\ref{tab:train}. After having trained the quantum circuit, we retrain the classical neural network with a larger number of shots $|\mathcal{D}_{L,\textrm{re}}^{(m)}|\gtrsim |\mathcal{D}_{L}^{(m)}|$ per state. This prevents overfitting at the expense of a moderate contribution $ \sum_{L=0,1} \sum_{m=1}^{M_L}|\mathcal{D}_{L,\textrm{re}}^{(m)}|$ to the total measurement budget, as this retraining is performed only once. 

\begin{figure}[t]
    \centering
    \includegraphics[width=\linewidth]{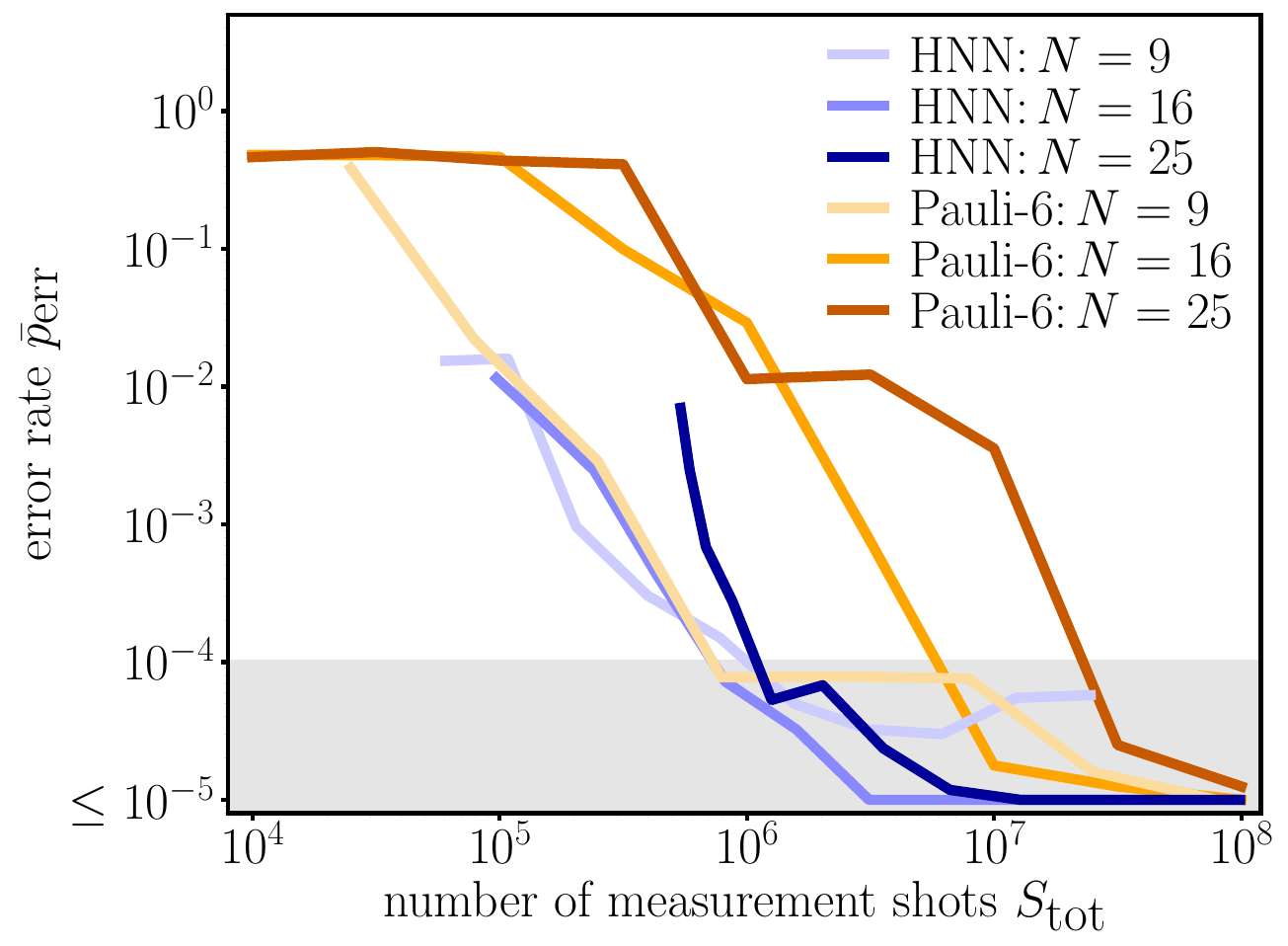}
    \caption{Sample complexity of learning the surface-code phase. The multi-shot error rate $\bar{p}_{\rm err}$ as a function of the total number $S_{\rm tot}$ of measurement shots used in training of the hybrid neural network (HNN) for $N=9$ and $S_{\rm inf} = 30$ (light blue line), $N=16$ and $S_{\rm inf} = 15$ (medium blue line), and $N=25$ and $S_{\rm inf} = 15$ (dark blue line), compared to the multi-shot error rate of the classical neural network trained on Pauli-6 POVM measurements for $N=9$ and $S_{\rm inf} = 200$ (light orange line), $N=16$ and $S_{\rm inf} = 100$ (medium orange line), and $N=25$ and $S_{\rm inf} = 75$ (dark orange line).}
    \label{fig:complexity}
\end{figure}

\begin{figure}[t]
    \centering
    \includegraphics[width=\linewidth]{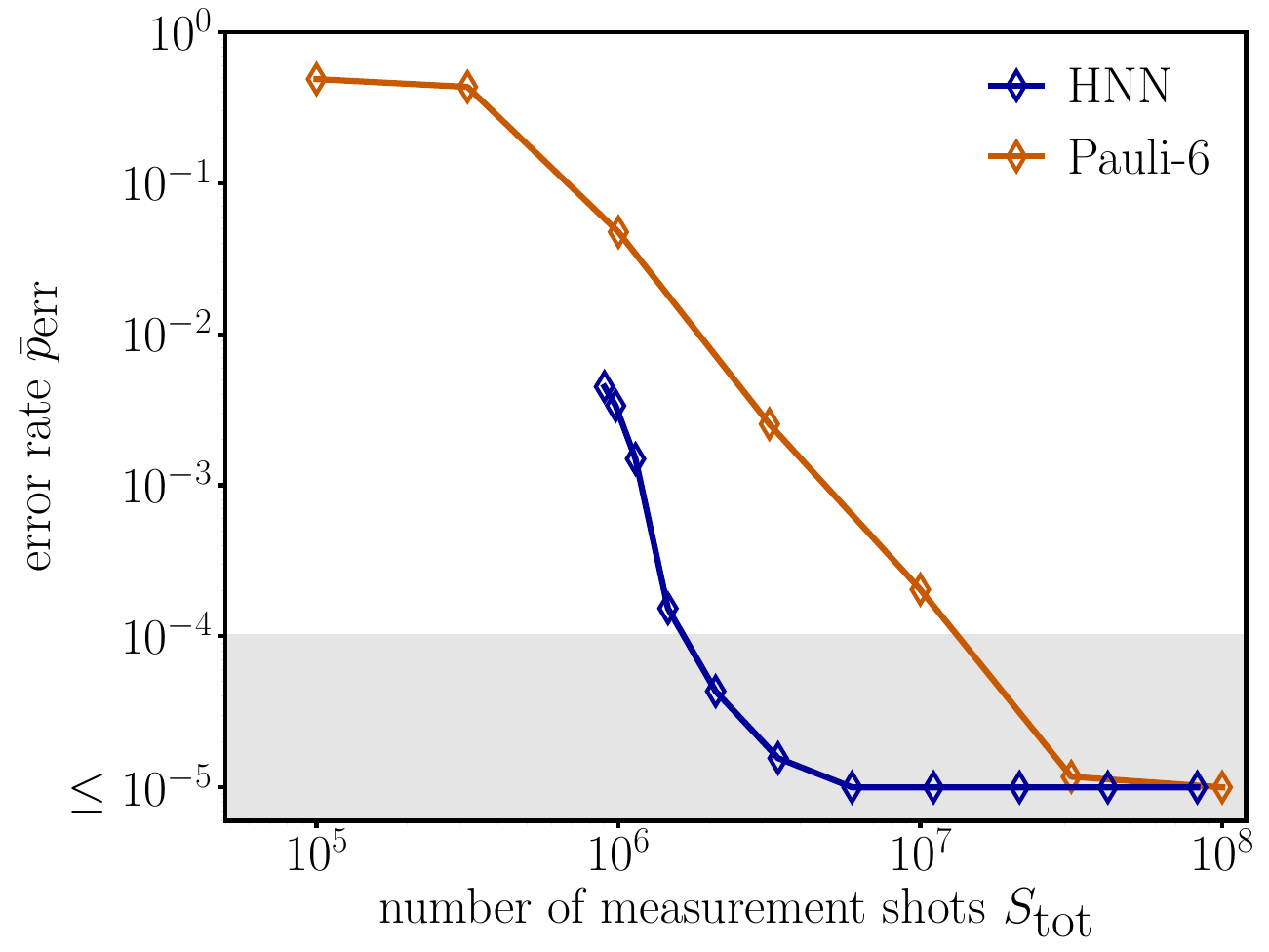}
    \caption{Sample complexity of learning the surface-code phase and the symmetry-enriched topological phase. The multi-shot error rate $\bar{p}_{\rm err}$ as a function of the total number $S_{\rm tot}$ of measurement shots used in training of the hybrid neural network (HNN) for $S_{\rm inf} = 25$ (blue diamonds), compared to the multi-shot error rate of the classical neural network trained on Pauli-6 POVM measurements for $S_{\rm inf} = 75$ (orange diamonds).}
    \label{fig:tns_complexity}
\end{figure}

To investigate the training sample complexity for the \emph{surface code in a magnetic field}, we vary the number of measurements used during training for different system sizes $N=9,16,25$. In particular, we set $|\mathcal{D}_1^{(m)}|=S_{\rm tr} $  equal for all $m=1,2,3$, draw  $3S_{\rm tr} $ bit strings from the 1-design and sweep $S_{\rm tr}$. To evaluate the performance of the trained hybrid neural network, we then determine the multi-shot error rate $\bar{p}_{\rm err} = (\bar{p}_{\rm fp} + \bar{p}_{\rm fn})/2$ using the topological ground states for $h=0,0.05,0.1,0.15,0.2,0.25$ and $10^4$ random product states. For $N=25$, numerically simulating the processing of product states in the hybrid neural network is too computationally demanding and we thus reduce their number to $10^3$. We plot in Fig.~\ref{fig:complexity} the multi-shot error rate $\bar{p}_{\rm err}$ averaged over 20 random initializations as a function of the total sample budget $S_{\rm tot}$  (blue lines). Note that we use sufficiently many shots $S_{\rm inf}$ in inference, as indicated in the caption of Fig.~\ref{fig:complexity}, such that errors stem from imperfect training. One can see that, for all system sizes, the hybrid neural network achieves $\bar{p}_{\rm err}<0.01\%$ with $S_{\rm tot} \approx 10^6$ training shots.

For the classical benchmark network, we first acquire training datasets $\mathcal{D}_L^{(m)}$ using the Pauli-6 POVM measurements and, in contrast to the hybrid neural network, do not perform any further measurements during training. Therefore, the total number of measurement shots is $S_{\rm tot} = \sum_{L=0,1} \sum_{m=1}^{M_L}|\mathcal{D}_L^{(m)}|$, where we set $|\mathcal{D}_1^{(m)}|=S_{\rm tr} $ equal for all $m=1,2,3$, draw  $3S_{\rm tr}$ POVM measurement outcomes from the 1-design and vary $S_{\rm tr}$. Note that, for $N=9,16$, three hidden layers with $n=256$ nodes are sufficient for the classical benchmark network, whereas five hidden layers are used for $N=25$.  We plot in Fig.~\ref{fig:complexity} the multi-shot error rate $\bar{p}_{\rm err}$ averaged over 20 random initializations of the classical benchmark network as a function of the total sample budget $S_{\rm tot}$  (orange lines). In contrast to the hybrid neural network, the number of training shots required for $\bar{p}_{\rm err}<0.01\%$ increases with system size. Moreover, for $N=25$ qubits, the classical benchmark network needs more than an order of magnitude more shots than the hybrid neural network.

We observe a similar reduction of training sample complexity for the dataset comprising tensor-network states spanning the \emph{transition between the surface-code phase and the symmetry-enriched topological phase}. In Fig.~\ref{fig:tns_complexity}, we plot the multi-shot error rate $\bar{p}_{\rm err}$ averaged over 10 random initializations of the hybrid neural network (blue line) and the classical benchmark network (orange line) for $N=24$ qubits. The error rate is evaluated on the test set and we exclude states for $|g|<0.15$ lying close to the phase transition to focus on the sample complexity of recognizing typical states well within each phase. The hybrid neural network achieves $\bar{p}_{\rm err}<0.01\%$ with approximately $S_{\rm tot} \approx 10^6$ training shots, whereas the classical benchmark network requires $S_{\rm tot}\approx 10^7$ training shots.

 In summary, the hybrid neural network reduces the shot number required to achieve $\bar{p}_{\rm fp},\,\bar{p}_{\rm fn}<0.01\%$ in inference compared to the classical benchmark network by approximately one order of magnitude for both datasets. Furthermore, its training sample complexity is approximately one order of magnitude smaller than the training sample complexity of the classical benchmark network.

\section{Conclusions}

We designed a hybrid neural network that consists of a classical neural network, which approximates an optimal classifier limited only by the statistical distance between measured data, and a parameterized quantum circuit, which variationally maximizes this distance to facilitate topological phase classification.
This hybrid neural network reduces both inference and training sample complexities by approximately one order of magnitude compared to a classical feedforward neural network trained on randomized Pauli-6 POVM measurements. This reduction is observed when distinguishing the topological phase of the surface code from both the ensemble of product states forming a 1-design and the symmetry-enriched topological phase. By introducing two datasets that are more efficiently classified when processed on a quantum computer, we thus address the lack of classically hard quantum data in quantum machine learning, which has become apparent as existing approaches often rely on locally easy data \cite{bermejo2026}.

Classifying a featureless ensemble of states, as in our first dataset, provides a simple test of whether a model learns genuine features of a target topological phase or only detects the absence of local features. Here, the 1-design of product states is a sufficient candidate, as both models learn expectation values of observables, which are linear functions of input quantum states. For nonlinear models such as shadow kernels \cite{huang2022} that produce $t$-th–order polynomials of input quantum states, the corresponding featureless ensemble should form a $t$-design. Such a nonlinear classical post-processing procedure is an interesting future direction for extending the hybrid neural network to estimate nonlinear properties of quantum states, such as the topological entanglement entropy \cite{kitaev2006}.

Detecting genuine features of topological phases is required to distinguish them from one another, as we demonstrated for our second dataset comprising the surface-code phase and the symmetry-enriched topological phase. Therefore, the hybrid neural network introduced here enables the efficient characterization and investigation of complex topological phases in two-dimensional systems, which are a subject of active research \cite{haller2023,liu2024}. For implementations on near-term quantum computers, robustness against decoherence, gate infidelities and state-preparation-and-measurement errors is essential. The companion paper \cite{scarato2026} demonstrates robust phase classification under such conditions on a superconducting quantum processor, and investigates the effects of state-preparation errors in detail, supporting the experimental viability of this approach.

By showing their trainability, we extend hybrid quantum-classical approaches for phase recognition to situations where we lack prior knowledge for explicit quantum circuit constructions \cite{herrmann2022,sander2025,zapletal2024}. In this work, the quantum circuit is constructed by inverting and parameterizing the preparation circuit for the unperturbed surface code \cite{satzinger2021}, whose depth scales as $\mathcal{O}(\sqrt{N})$. Developing less problem-specific and more systematic approaches to designing circuits with more favorable depth scaling while maintaining comparable performance is an important direction for future work.

\section{ACKNOWLEDGMENTS}
We thank Andreas Wallraff, Christoph Bruder, Niels Lörch and Julian Arnold for helpful discussions and feedback.
This work is part of the Munich Quantum Valley, which is supported by the Bavarian state government with funds from the Hightech Agenda Bayern Plus, and it was supported by the EU program HORIZON-MSCA-2022-PF Project No. 101108476 HyNNet NISQ. C.S., C.H. and J.K. acknowledge financial support by ETH Zurich.

\appendix

\section{Parameterized quantum circuit}\label{sec:circ}

\begin{figure}[b]
    \centering
    \includegraphics[width=0.8\columnwidth]{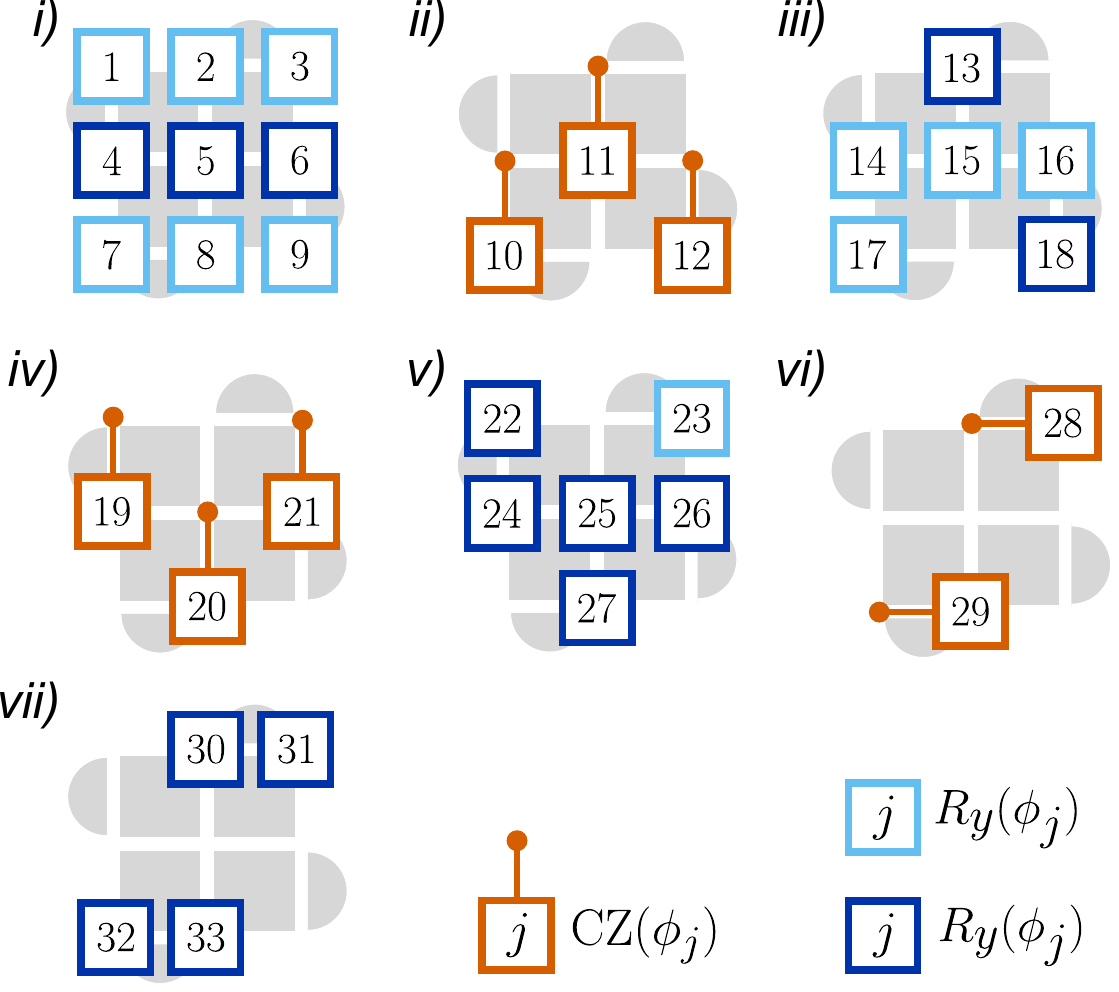}
    \caption{Parameterized quantum circuit for $N=9$ comprising layers i)-vii) of single-qubit gates $R_y(\phi_j) = \exp(-i\phi_j Y/2)$ (blue) and two-qubit gates $\textrm{CZ}(\phi_j)$ (orange), drawn on the surface-code lattice corresponding to Fig.~\ref{fig:hnn}a.}
    \label{fig:AppC_circuit9}
\end{figure}
\begin{figure*}[t]
    \centering
    \includegraphics[width=0.9\textwidth]{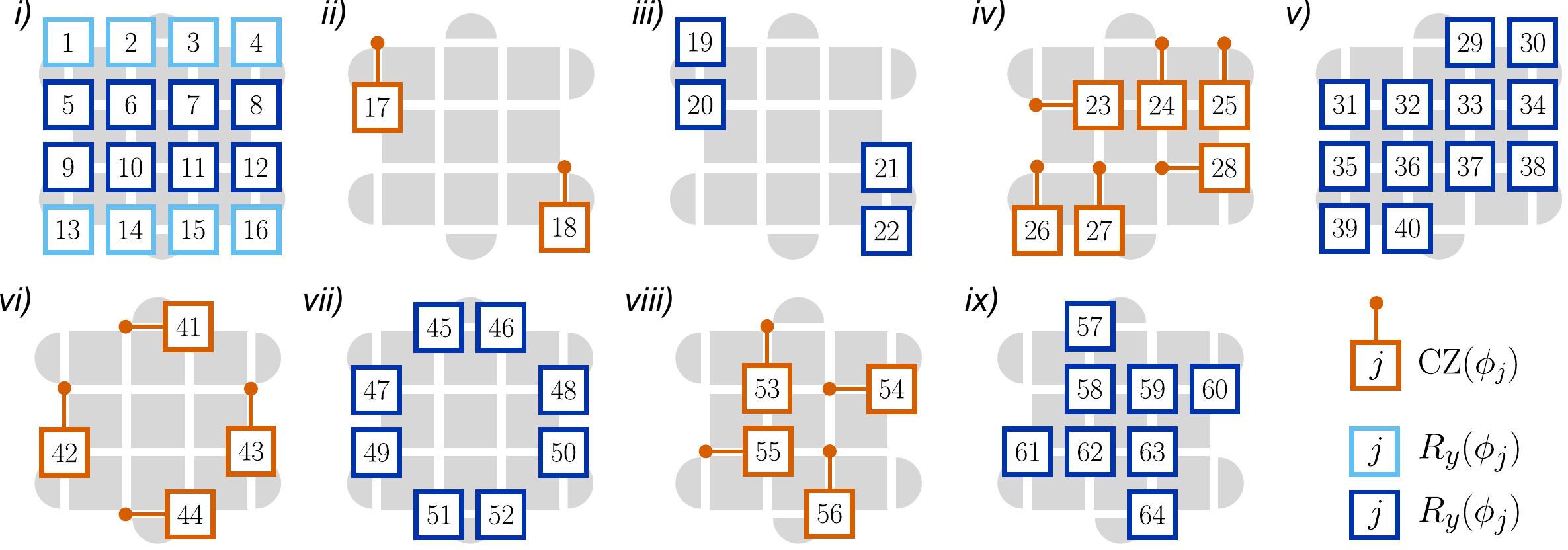}
    \caption{Parameterized quantum circuit for $N=16$ with layers i)-ix) of single-qubit gates (blue) and two-qubit gates (orange), similar to Fig.~\ref{fig:AppC_circuit9}.}
    \label{fig:AppC_circuit16}
\end{figure*}
\begin{figure*}[t]
    \centering
    \includegraphics[width=\textwidth]{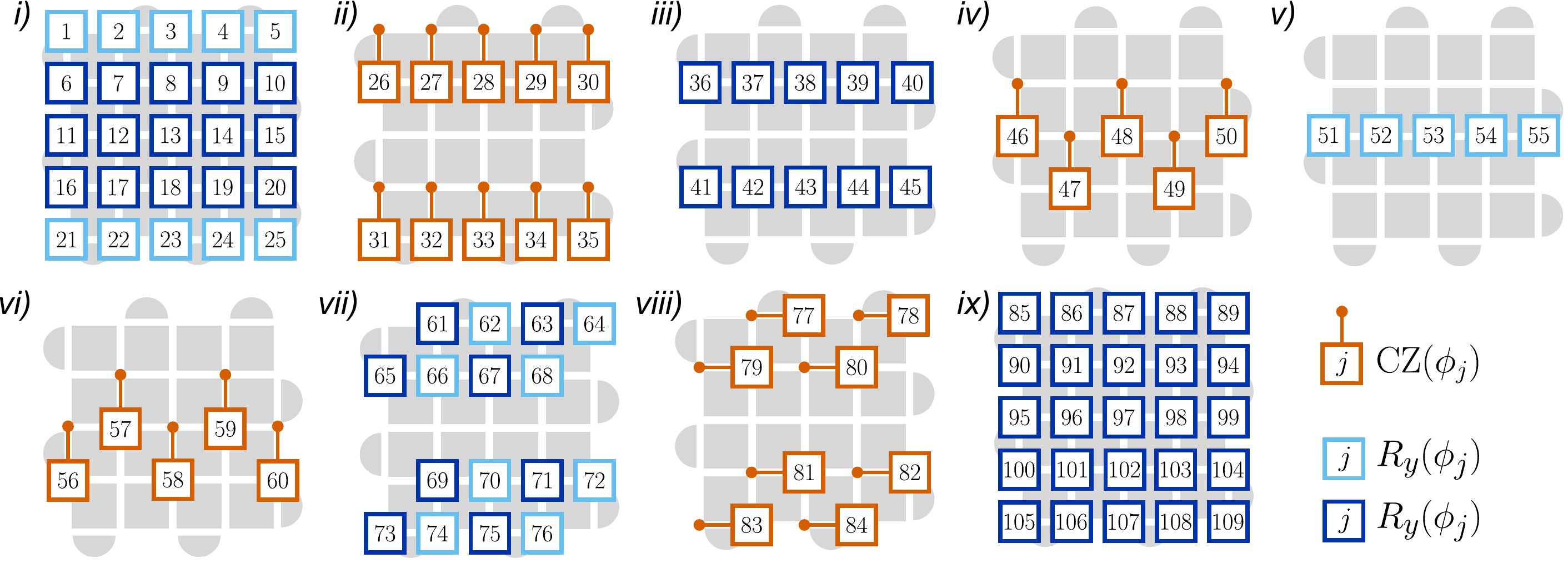}
    \caption{Parameterized quantum circuit for $N=25$ with layers i)-ix) of single-qubit gates (blue) and two-qubit gates (orange), similar to Fig.~\ref{fig:AppC_circuit9}.}
    \label{fig:AppC_circuit25}
\end{figure*}

In this appendix, we describe the parameterized quantum circuit and its construction. For the surface code with $N=9$ and $N=25$ qubits depicted in Fig.~\ref{fig:hnn}a, the circuit is based on the preparation circuit \cite{satzinger2021}, compiled into the native gates $R_y(\pm\frac{\pi}{2})$ and  $\textrm{CZ}(\pi)$ of the superconducting processor \cite{scarato2025}. For $N=16$, we start from the surface-code preparation circuit introduced in Ref.~\cite{scarato2026}.

To construct the parameterized quantum circuit depicted in Figs.~\ref{fig:AppC_circuit9}, \ref{fig:AppC_circuit16} and \ref{fig:AppC_circuit25} for $N=9$, $N=16$ and $N=25$, respectively, we invert the corresponding preparation circuit, and replace the gates $R_y(\pm\frac{\pi}{2})$ and $\textrm{CZ}(\pi)$ with gates $R_y(\phi_j)$ and $\textrm{CZ}(\phi_j)$ parameterized by angles and phases $\phi_j$, which are chosen independently. To enhance the expressivity of the circuit, we include additional parameterized single-qubit rotations $R_y(\phi_j)$, shown in light blue in Figs.~\ref{fig:AppC_circuit9}, \ref{fig:AppC_circuit16} and \ref{fig:AppC_circuit25}, ensuring that an $R_y(\phi_j)$ gate is applied to each qubit before every $\textrm{CZ}(\phi_j)$ gate and measurement. This results in $K=33,64,109$ parameterized gates for $N=9,16,25$, respectively. By construction, the parameterized quantum circuit implements the inverse of the surface code preparation circuit ($h=0$) for the parameters $\phi_{\rm inv}$ stated in Tabs.~\ref{tab:param}a, \ref{tab:param}b and \ref{tab:param}c for $N=9$, $N=16$ and $N = 25$, respectively.

\begin{figure*}[t]
    \centering
    \includegraphics[width=\textwidth]{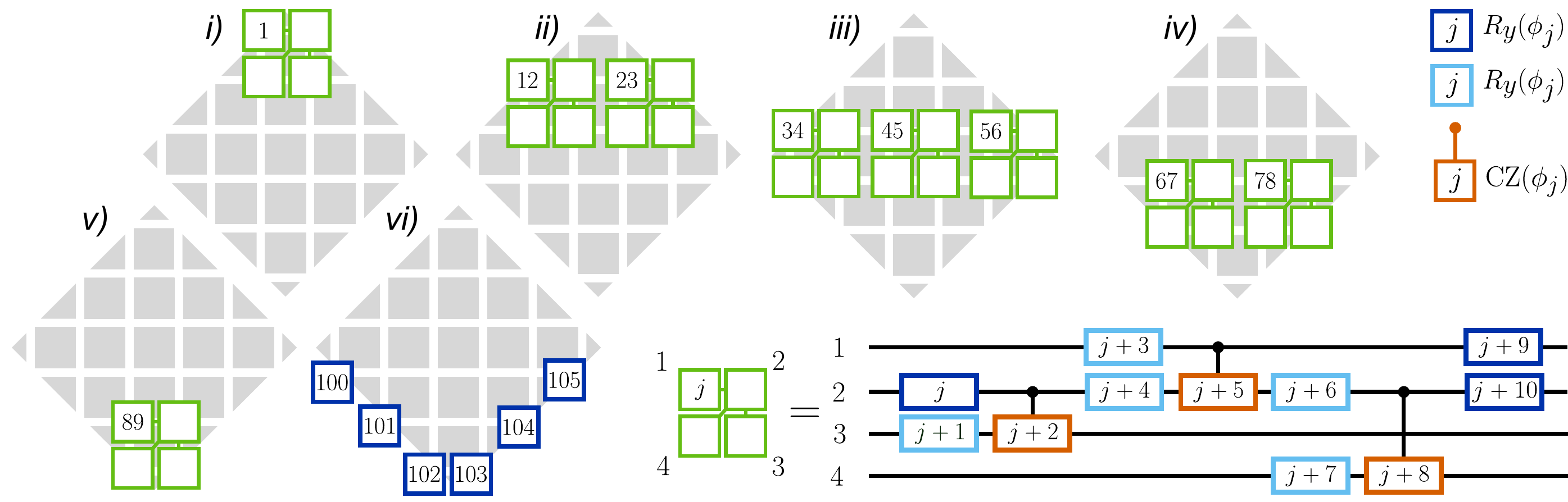}
    \caption{Parameterized quantum circuit for $N=24$ comprising layers i)-v) of single qubit gates $R_y(\phi_j) = \exp(-i\phi_j Y/2)$ (blue) and two-qubit gates $\textrm{CZ}(\phi_j)$ (orange) and layer vi) of single-qubit gates $R_y(\phi_j)$ (blue), drawn on the surface-code lattice with the open boundary conditions shown in Fig.~\ref{fig:rotated_SC}.}
    \label{fig:AppC_circuit24}
\end{figure*}

Analogously, we design the parameterized quantum circuit for the surface code with $N=24$ depicted in Fig.~\ref{fig:rotated_SC}. The starting point for this construction is the tensor-network preparation circuit introduced in Ref.~\cite{liu2024}, where we specifically consider the circuit for $g=1$ that prepares the surface-code ground state. The construction yields the circuit depicted in Fig.~\ref{fig:AppC_circuit24} with $K=105$ parameters. This circuit implements the inverse of the preparation circuit for the parameter values
\begin{align}
    \phi_{j + 1,\textrm{inv}},\phi_{j+3,\textrm{inv}},\phi_{j+4,\textrm{inv}}, \phi_{j+6,\textrm{inv}} ,\phi_{j+7,\textrm{inv}}  &= 0\\  
    \phi_{j + 10,\textrm{inv}}  &= \frac{\pi}{2}\\ 
    \phi_{j + 2,\textrm{inv}},\phi_{j+5,\textrm{inv}},\phi_{j+8,\textrm{inv}}  &= \pi\\
    \phi_{j,\textrm{inv}},\phi_{j+9,\textrm{inv}} &= \frac{3\pi}{2}          
\end{align}
for $j=1,12,23,34,45,56,67,78,89$ and 
\begin{equation}
\phi_{100,\textrm{inv}},\phi_{101,\textrm{inv}},\ldots,\phi_{105,\textrm{inv}} = \frac{3\pi}{2}.
\end{equation}

\begin{table}[b]
    \centering
    
    \textbf{(a) $N=9$}
    
    \vspace{0.3em}
    
    \begin{tabular}{c|c}
         $\phi_{j,\textrm{inv}}$ & index $j$ \\ \hline 
         $0$ & $1,2,3,7,8,9,14,15,16,17,23$ \\
         $\frac{\pi}{2}$ & $24,25,26,30,33$ \\
         $\pi$ & $10,11,12,19,20,21,28,29$ \\
         $\frac{3 \pi}{2}$ & $4,5,6,13,18,22,27,31,32$ \\
    \end{tabular}

    \vspace{0.3em}

    \textbf{(b) $N=16$}
    
    \vspace{0.3em}

    \begin{tabular}{c|c}
         $\phi_{j,\textrm{inv}}$ &  index $j$ \\ \hline 
         $0$ & $1, 2, 3, 4, 13, 14, 15, 16$ \\
         $\frac{\pi}{2}$ & $20, 21, 32, 33, \ldots, 37, 46, 47, 50, 51, 57, 60, 61, 64$ \\
         $\pi$ & $17, 18, 23, 24, \ldots, 28, 41, 42, 43, 44, 53, 54, 55, 56$ \\
         \multirow{2}{*}{$\frac{3 \pi}{2}$} & $5, 6, \ldots, 12, 19, 22, 29, 30, 31, 38, 39, 40, 45, 48, 49, 52,$\\
         &$ 58, 59, 62, 63$ \\
    \end{tabular}

    \vspace{0.3em}

    \textbf{(c) $N=25$}
    
    \vspace{0.3em}    

    \begin{tabular}{c|c}
         $\phi_{j,\textrm{inv}}$&  index $j$ \\ \hline 
         \multirow{2}{*}{$0$} & $1, 2, \ldots, 5, 21, 22, \ldots, 25, 51, 52, \ldots, 55, 62, 64, 66, 68,$\\
         &$70, 72, 74, 76$ \\
         \multirow{2}{*}{$\frac{\pi}{2}$} & $36, 37, \ldots, 45, 86, 88, 90, 92, 95, 96, 97, 98, 99, 101, 103,$\\
         & $ 105, 107$ \\
         $\pi$ & $26, 27, \ldots, 35, 46, 47, \ldots, 50, 56, 57, \ldots, 60, 77,$\\
         &$78, \ldots, 84$ \\
         \multirow{2}{*}{$\frac{3 \pi}{2}$} & $6, 7, \ldots, 20, 61, 63, 65, 67, 69, 71, 73, 75, 85, 87, 89, 91,$\\
         & $93, 94, 100, 102, 104, 106, 108, 109$ \\
    \end{tabular}
    
    \caption{Parameter values $\phi_{\rm inv}$ for which the quantum circuits in Figs.~\ref{fig:AppC_circuit9}, \ref{fig:AppC_circuit16} and \ref{fig:AppC_circuit25} implement the inverse surface-code preparation circuit for (a) $N=9$, (b) $N=16$ and (c) $N=25$, respectively.}
    \label{tab:param}
\end{table}

\section{Hybrid neural network training}\label{sec:train}

Here, we describe the details of the hybrid neural network training using Algorithm \ref{tab:train}. In the inner loop, we train the feedforward neural network on datasets $\mathcal{D}_L^{(m)}$ obtained by measuring the training states $\ket{\psi_L^{(m)}}$ processed in the parameterized quantum circuit $U(\phi^{(k)})$. The input nodes receive the measurement outcomes $x_i$, standardized as $x_i \mapsto (x_i - \bar{x}_i)/s_i$, where $\bar{x}_i$ and $s_i$ are the mean and standard deviation over training datasets $\mathcal{D}_L^{(m)}$ for all $L$ and $m$. We build the feedforward neural network in Keras within TensorFlow and train it using backpropagation \cite{baydin2018} with gradient descent using the Adam optimizer \cite{kingma2017}.

Additionally, we train the feedforward neural network on datasets of bit strings obtained for stochastic perturbations of the parameters $\phi^{(k)}+\delta\phi^{(k)}$, which are referred to as a population. In each iteration $k$, these stochastic perturbations $\delta\phi^{(k)}$ are drawn from a normal distribution with zero mean and standard deviation $s_{\rm pop}$. In the outer loop of Algorithm \ref{tab:train}, we use the evolution strategy optimizer to estimate the gradient of the residual cost $\textrm{grad}_{\phi}C_{\rm res}$ from the $k$th population and update the circuit parameters $\phi^{(k+1)} \gets \phi^{(k)} -\alpha\, \textrm{grad}_{\phi} C^{(k)}_{\rm res}$.

For the surface code in a magnetic field \eqref{eq:field}, the training dataset ($L=1$) comprises ground states of the topological phase, where the phase boundary with the trivial phase is located at $h_c=0.17,0.28,0.3$ for $N=9,16,25$, respectively. In particular, we choose topological ground states $\ket{\Psi_1^{(m)}}$, for $m=1,2,3$, at field strengths $h=0,0.05,0.15$ for $N=9$ and at $h=0,0.1,0.2$ for $N=16, 25$.

\begin{figure}[t]
    \centering
    \includegraphics[width=\linewidth]{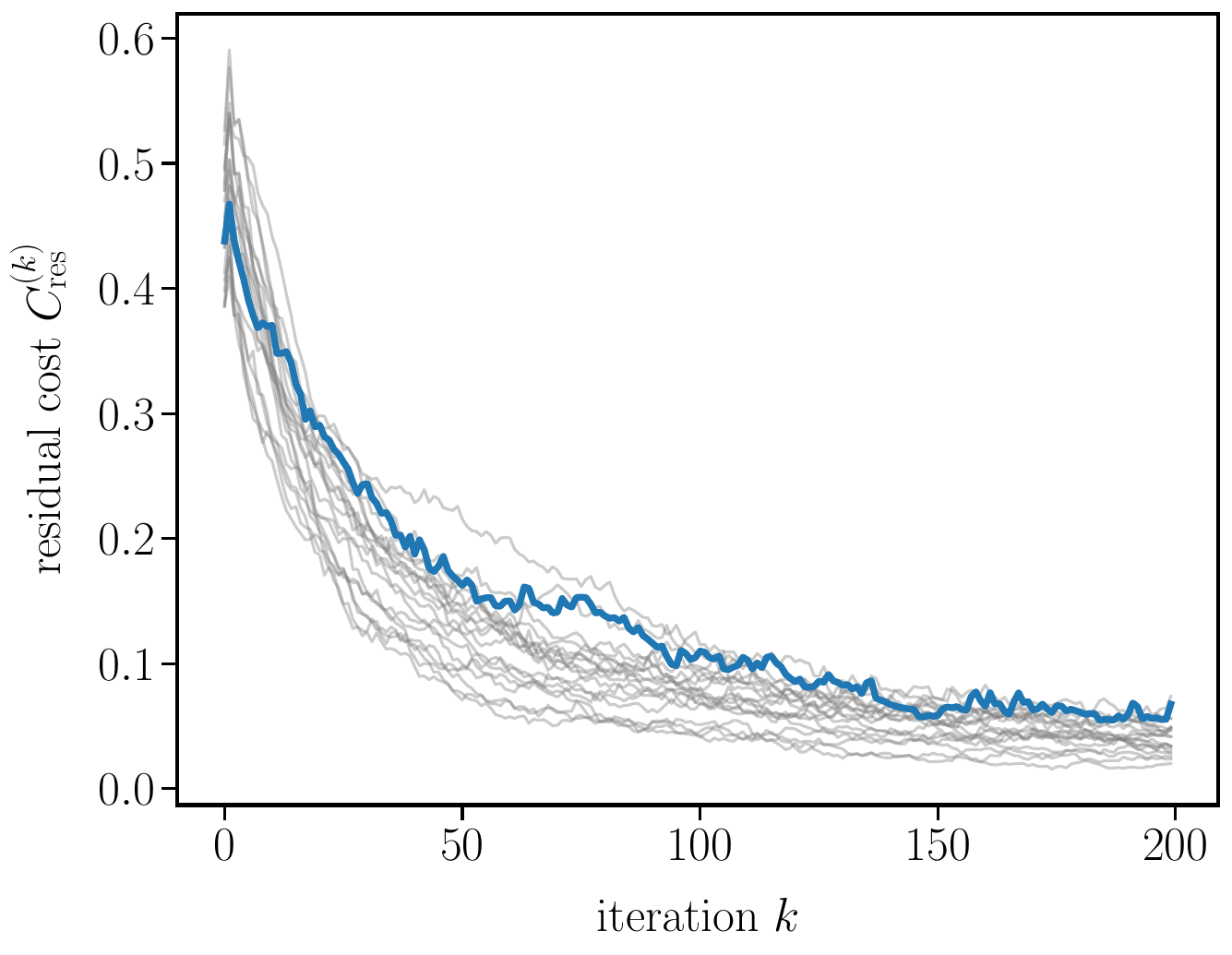}
    \caption{Residual cost $C^{(k)}_{\rm res}$ at each iteration $k$ of the outer optimization loop for $N=25$ and 20 random initializations of the parameters $\phi^{(1)}$.}
    \label{fig:training_traces}
\end{figure}

In Fig.~\ref{fig:training_traces}, we plot the residual cost $C_{\rm res}$ for $N=25$ at each iteration $k$ of the outer optimization loop for 20 random initializations of the parameters $\phi^{(1)}$. The optimization converges within 200 iterations, reaching the residual cost $C_{\rm res} = 0.04 \pm 0.01$, which exhibits a variation across the random initializations due to local minima in the cost landscape. The circuit parameters trained for the blue trajectory are used for multi-shot inference in Fig.~\ref{fig:inference} and the recognition of the topologically trivial magnetic phase in Fig.~\ref{fig:field}.

\subsection{Classical neural network tuning}

To tune the hyperparameters of the feedforward neural network, we train it on datasets $\mathcal{D}_L^{(m)}$ measured for 10 different instances of random initial circuit parameters $\phi^{(1)}$, which are harder to classify than measurement outcomes obtained in subsequent iterations $k>1$. By training the quantum circuit, we reduce the statistical overlap between the datasets $\mathcal{D}_1^{(m)}$ and $\mathcal{D}_0^{(m)}$ and, thereby, facilitate their classification.

The training converges within $N_{\rm epoch} = 1000$ epochs for the learning rate $\lambda = 0.01$ of the Adam optimizer. Note that $\lambda$ is different from the learning rate $\alpha$ of the evolution strategy optimizer. For smaller learning rates,  the required number of epochs increases without any improvement in the residual cost. On the other hand, increasing the learning rate hinders training and leads to a larger residual cost. Next, by tuning the number of hidden nodes $n$, we find that the network reaches the minimal residual cost $C_{\rm res}^{\rm (val)}$ evaluated on validation datasets $\mathcal{D}^{(m)}_{L,\textrm{val}}$, which are unseen during training, with $n=9,32,128$ for $N=9,16,25$, respectively. Finally, the number of training shots $S_{\rm tr}=|\mathcal{D}_1^{(m)}| = 2^{11},2^{13},2^{16}$ per topological state is sufficient for $N=9,16,25$, respectively, to reach the difference between the validation and training costs $C_{\rm res}^{\rm (val)} -C_{\rm res}^{\rm (tr)}$ comparable or smaller than the difference $C_{\rm res}^{\rm (tr)} - C_{\rm opt}$ from the global optimum $ C_{\rm opt}$ given by Eq.~\eqref{eq:bxe_opt}. From the 1-design, we draw $3S_{\rm tr}$ shots.

In the subsequent optimization iterations $k>1$, the feedforward neural network is initialized with the weights and biases $w^{(k-1)}$ trained in the previous iteration $k-1$ for the current circuit parameters $\phi^{(k)}$ and the full population $\phi^{(k)}+\delta\phi^{(k)}$. Since the circuit parameters are updated by small steps $\phi^{(k+1)} \gets \phi^{(k)} -\alpha\, \textrm{grad}_{\phi} C^{(k)}_{\rm res}$, the bit string datasets $\mathcal{D}_L^{(m)}$ remain similar in successive iterations. As a result, the training converges for a substantially smaller number of epochs  $N_{\rm epoch}$ than in the first iteration. In particular, we use $N_{\rm epoch}=100$ for $N=9,16$ and $N_{\rm epoch}=10$ for $N=25$.

For $N=9,16$, the cost difference $C_{\rm res}^{\rm (val)} -C_{\rm res}^{\rm (tr)}\sim0.01$ is small, showing that the network does not suffer from overfitting. A comparable cost difference $C_{\rm res}^{\rm (val)} -C_{\rm res}^{\rm (tr)}$ is retained for trained circuit parameters $\phi^*$. However, the cost difference is larger $C_{\rm res}^{\rm (val)} -C_{\rm res}^{\rm (tr)}\approx 0.04$ for $N=25$. While this does not hinder the training of the quantum circuit, it leads to classification errors for the trained network. To mitigate this overfitting, we use early stopping \cite{googfellow2016}. To this end, we retrain the feedforward neural network on datasets of bit strings $\mathcal{D}^{(m)}_{L,\textrm{re}}$ measured for the trained circuit parameters $\phi^*$. We split these datasets into training datasets $\mathcal{D}^{(m)}_{L,\textrm{tr}}$ and validation datasets $\mathcal{D}^{(m)}_{L,\textrm{val}}$ of size $|\mathcal{D}^{(m)}_{L,\textrm{tr}}|=4|\mathcal{D}^{(m)}_{L,\textrm{re}}|/5$ and $|\mathcal{D}^{(m)}_{L,\textrm{val}}|=|\mathcal{D}^{(m)}_{L,\textrm{re}}|/5$. After each epoch, we evaluate the cost on the validation datasets $\mathcal{D}^{(m)}_{L,\textrm{val}}$ and the retraining is terminated if the validation cost does not decrease for 20 epochs.

Analogously, we tune the hyperparameters for the training states spanning the phase transition between the surface-code phase and the symmetry-enriched topological phase: $N_{\rm epoch} = 1000$ in the first iteration, $N_{\rm epoch} = 30$ in all subsequent iterations,  $\lambda=0.002$, $n=128$, $S_{\rm tr} = 2^{16}$. We also retrain the feedforward network for the trained circuit parameters $\phi^*$ and use early stopping to mitigate overfitting.

\subsection{Evolution strategy tuning}
To isolate the effects of the evolution strategy hyperparameters on the quantum circuit training, we replace the feedforward neural network with the optimal analytical postprocessing procedure \cite{arnold2022}. To this end, we analytically determine the probability $y_{\rm opt}(x)$ for every possible measurement outcome $x$ from the full probability distributions $P_1(x)$ and $P_0(x)$ according to Eq.~\eqref{eq:yopt}. This eliminates imperfections arising from the training of the feedforward neural network and allows us to assess the impact of the hyperparameters on the circuit optimization in a controlled setting.

The standard deviation $s_{\rm pop}$ of the population largely affects the circuit optimization. To tune it, we set the learning rate $\alpha=s_{\rm pop}^2$ equal to the variance. For small $s_{\rm pop}$, the estimation of the residual cost gradient is very precise but the optimizer requires many iterations to converge due to small parameter updates as we use the correspondingly small learning rate $\alpha=s_{\rm pop}^2$. On the other hand, the estimation of the residual cost gradient is not precise enough for large $s_{\rm pop}$, hindering the circuit optimization. A good trade-off between the speed and precision of the optimization is achieved for $s_{\rm pop}/\pi=0.04$, where the optimizer reaches small cost values within $N_{\rm it}=200$ iterations. For learning rates smaller than $\alpha=s_{\rm pop}^2$, the optimization converges more slowly without any improvement in the achieved cost. On the other hand, the optimization is hindered for larger rates. The population size $N_{\rm pop}=10$ is sufficient to train the circuit parameters, as increasing $N_{\rm pop}$ further does not lead to any significant reduction of the cost. 

For these hyperparameters, the evolution strategy optimizer performs well for all training states. This includes the tensor-network states spanning the surface-code phase and the symmetry-enriched topological phase, as well as for the surface code in a magnetic field for all system sizes. This consistent performance can be attributed to the comparable number of parameters in the quantum circuit for all these datasets.

\section{Topologically trivial magnetic phase}\label{sec:trivial}
\begin{figure}
    \centering
    \includegraphics[width=\linewidth]{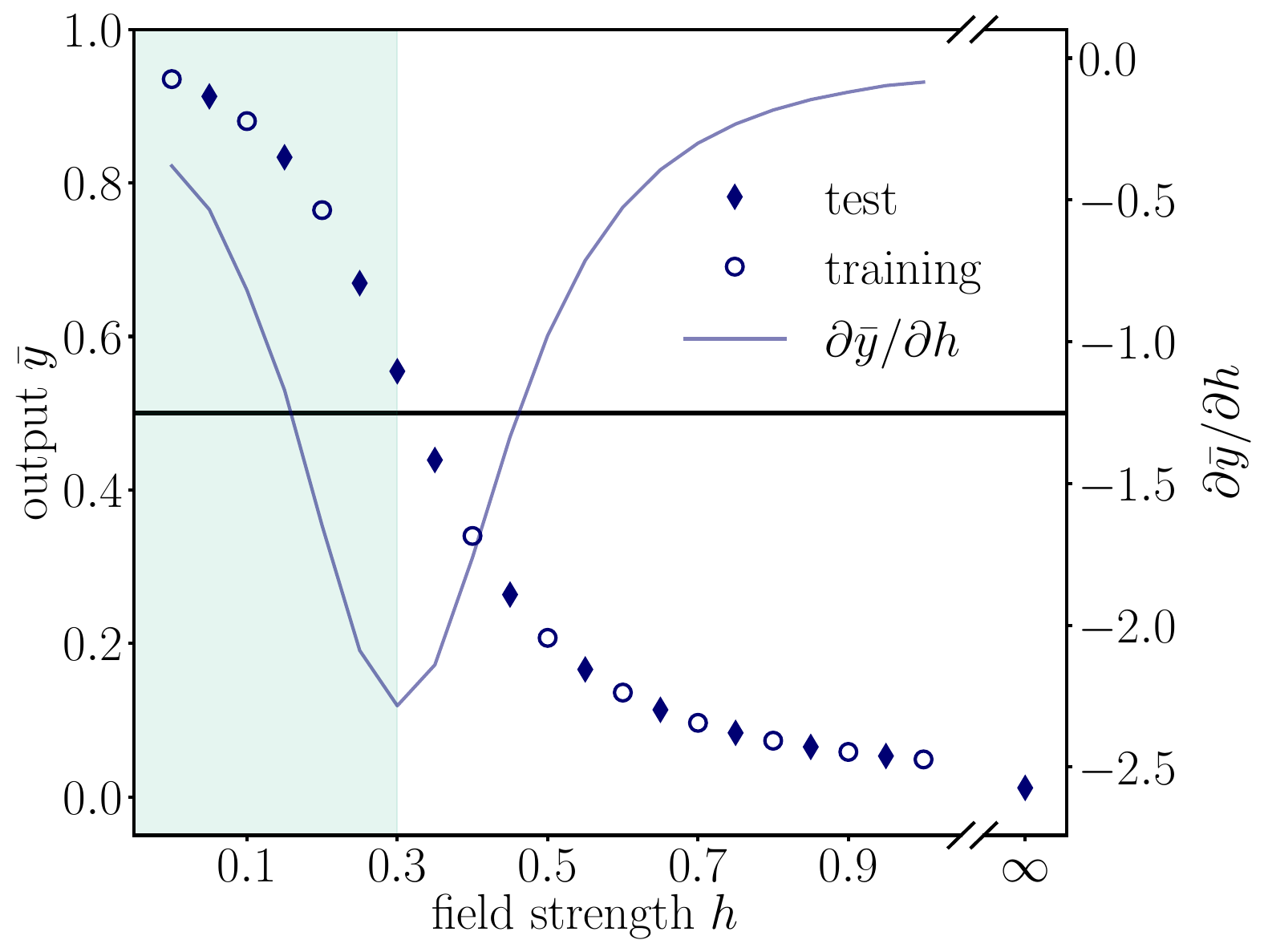}
    \caption{Recognizing the topological phase and the trivial phase of the surface code using a retrained hybrid neural network. Mean output $\bar{y}$ of the hybrid neural network as a function of the magnetic field strength $h$ for ground states from the training set (circles) and test set (diamonds). The shaded region corresponds to the topological phase and the decision threshold $t=0.5$ is shown as the black line. We also show a rescaled slope of the output, estimated using the Savitzky-Golay filter with window length $\omega=5$ and polynomial order $o=3$.}
    \label{fig:field}
\end{figure}

In this appendix, we investigate the classification of the topologically trivial ground states of the surface code for $h>h_c$ by adopting an approach similar to quantum-to-classical transfer learning introduced in Ref.~\cite{mari2020}. Since detecting the absence of topological features is sufficient to classify these trivial states, we use the quantum circuit $U(\phi^*)$, which was trained on the 1-design, to facilitate the measurement of these features. However, the probability distribution $P_{\rm trivial}(x)$ of the measurement outcomes $x$ for these trivial ground states is very different from the uniform distribution $P_0(x)$ of the 1-design, with the near-maximal total variation distance $\textrm{TV}[P_0,P_{\rm trivial}]=0.98$ and Jensen-Shannon divergence $D_{JS}[P_0,P_\mathrm{trivial}]=0.67$. Therefore, we retrain the weights and biases $w$ of the feedforward neural network to estimate the probability $1 - y_{\rm opt}(x)$ of measuring the outcomes $x$ for these trivial states. 

In doing so, we retrain the feedforward neural network for 1000 epochs using the topological ground states for $h=0,0.1,0.2$ and the trivial ground states for $h=0.4,0.5,0.6,0.7,0.8,0.9,1$ as training data with labels $L=1$ and $L=0$, respectively, with $|\mathcal{D}_L^{(m)}|=2^{16}$ bit strings measured for each state. For single-shot measurements, we find the false negative error rate $p_{\rm fn}=12.5 \%$ and the false positive error rate $p_{\rm fp}=10.2\%$. In Fig.~\ref{fig:field}, we plot the mean output $\bar{y}$ of the retrained hybrid neural network as a function of the field strength $h$, showing that it correctly classifies all topological and trivial ground states also for unseen field strengths during both the training of the quantum circuit and the retraining of the feedforward neural network.

\section{Symmetry-enriched topological phase}\label{sec:set_app}

\begin{figure}
    \centering
    \includegraphics[width=0.5\linewidth]{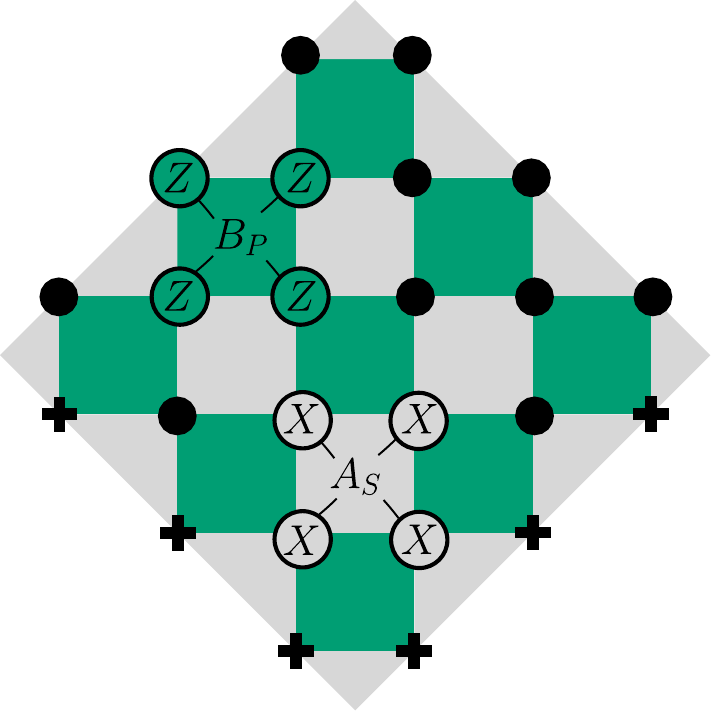}
    \caption{Surface code for $N=24$ qubits, whose ground state is the tensor-network state for $g=1$. Qubits are depicted as black points and crosses, and weight-four operators $A_s$ and $B_p$ are shown as gray and green squares, respectively. Weight-two and weight-three $A_s$ operators are depicted at the boundary as gray triangles.}
    \label{fig:rotated_SC}
\end{figure}

The tensor networks introduced in Ref.~\cite{liu2024} describe states of $N$ qubits located on a square lattice as shown for $N=24$ in Fig.~\ref{fig:rotated_SC}. To numerically simulate these states, we use the exact preparation circuits proposed in Ref.~\cite{liu2024}. In this preparation procedure, the qubits at the bottom boundary, depicted as black crosses, are initialized in the state $\ket{+}$ and the remaining qubits, depicted as black points, are initialized in the state $\ket{0}$. For $g=1$, the tensor-network state is the unique ground state of the surface code with open boundary conditions involving only weight-two and weight-three $A_s$ operators at the boundary (gray triangles in Fig.~\ref{fig:rotated_SC}), in contrast to Fig.~\ref{fig:hnn}a, where both $A_s$ and $B_p$ operators appear at the boundary. 

\begin{figure}
    \centering
    \includegraphics[width=\linewidth]{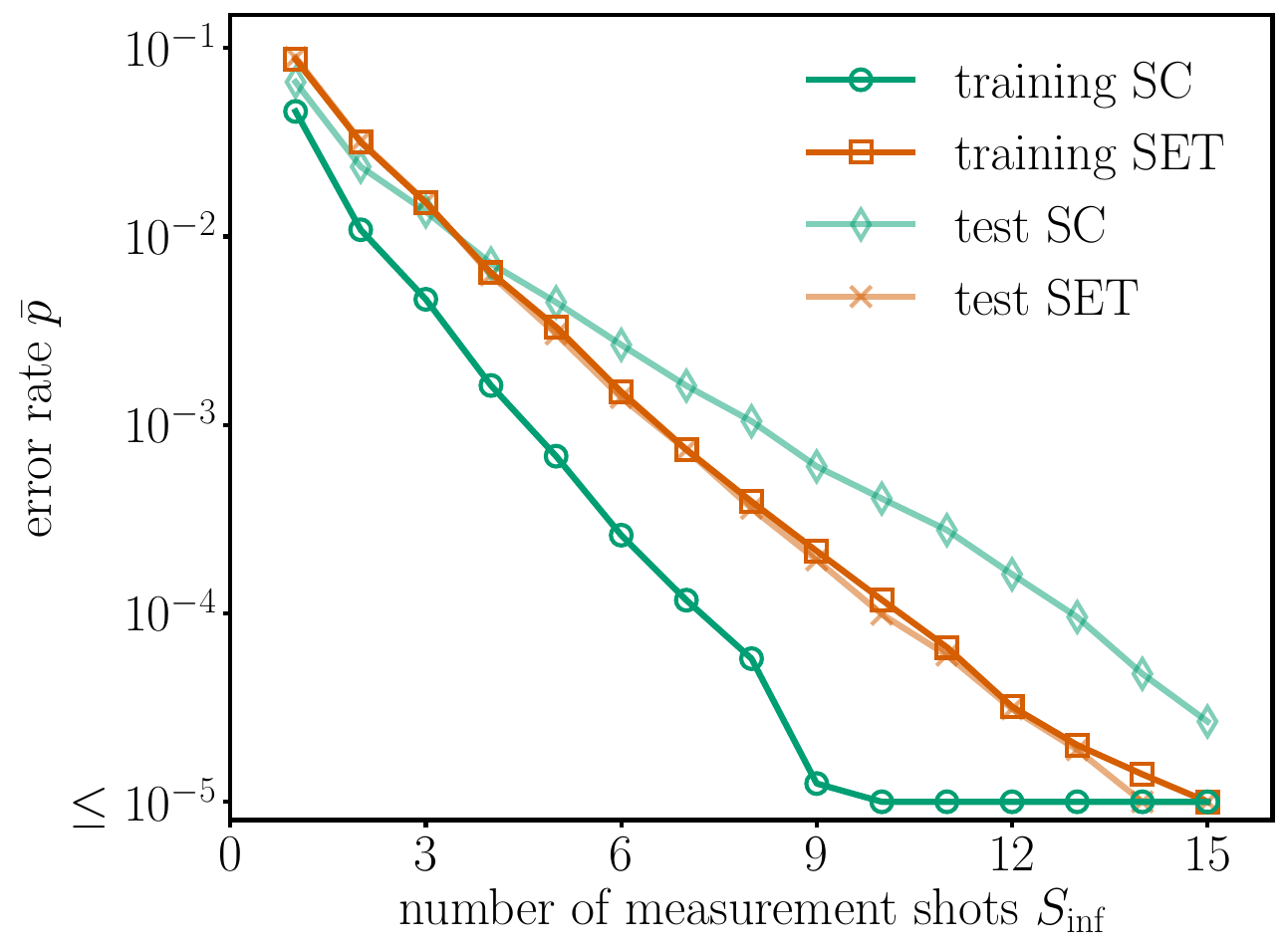}
    \caption{Multi-shot inference for recognizing the surface-code (SC) phase and the symmetry-enriched topological (SET) phase. The false negative error rate $\bar{p}_{\rm fn}$ for the training SC states (dark green circles) and the test SC states (light green diamonds), as well as the false positive error rate $\bar{p}_{\rm fp}$ for the training SET states (dark red squares) and the test SET states (light red crosses) as a function of the measurement shot number $S_{\rm inf}$.
    }
    \label{fig:tns_inference}
\end{figure}

Using the hybrid neural network trained in Sec.~\ref{sec:set}, we perform multi-shot inference and plot error rates for different tensor-network states as a function of the shot number $S_{\rm inf}=|\mathcal{D}^{(m)}_L|$ in Fig.~\ref{fig:tns_inference}. We can see that the multi-shot error rates rapidly decrease with $S_{\rm inf}$ for all states, including training states from the surface-code phase (dark green circles) and the symmetry-enriched topological phase (dark red squares), as well as test states from the surface-code phase (light green
diamonds) and the symmetry-enriched topological phase (light red crosses). For $S_{\rm inf}=13$, the hybrid neural network achieves near-perfect classification, with $\bar{p}_{\rm fp},\,\bar{p}_{\rm fn}<0.01\%$. In this plot, we exclude states for $|g|<0.15$ lying close to the phase transition to focus on the sample complexity of recognizing typical states well within each phase.

\section{Classical benchmark network}\label{sec:cnn}

\begin{figure}[t]
    \centering
	\hspace{0.2cm}
	\includegraphics[width=\linewidth]{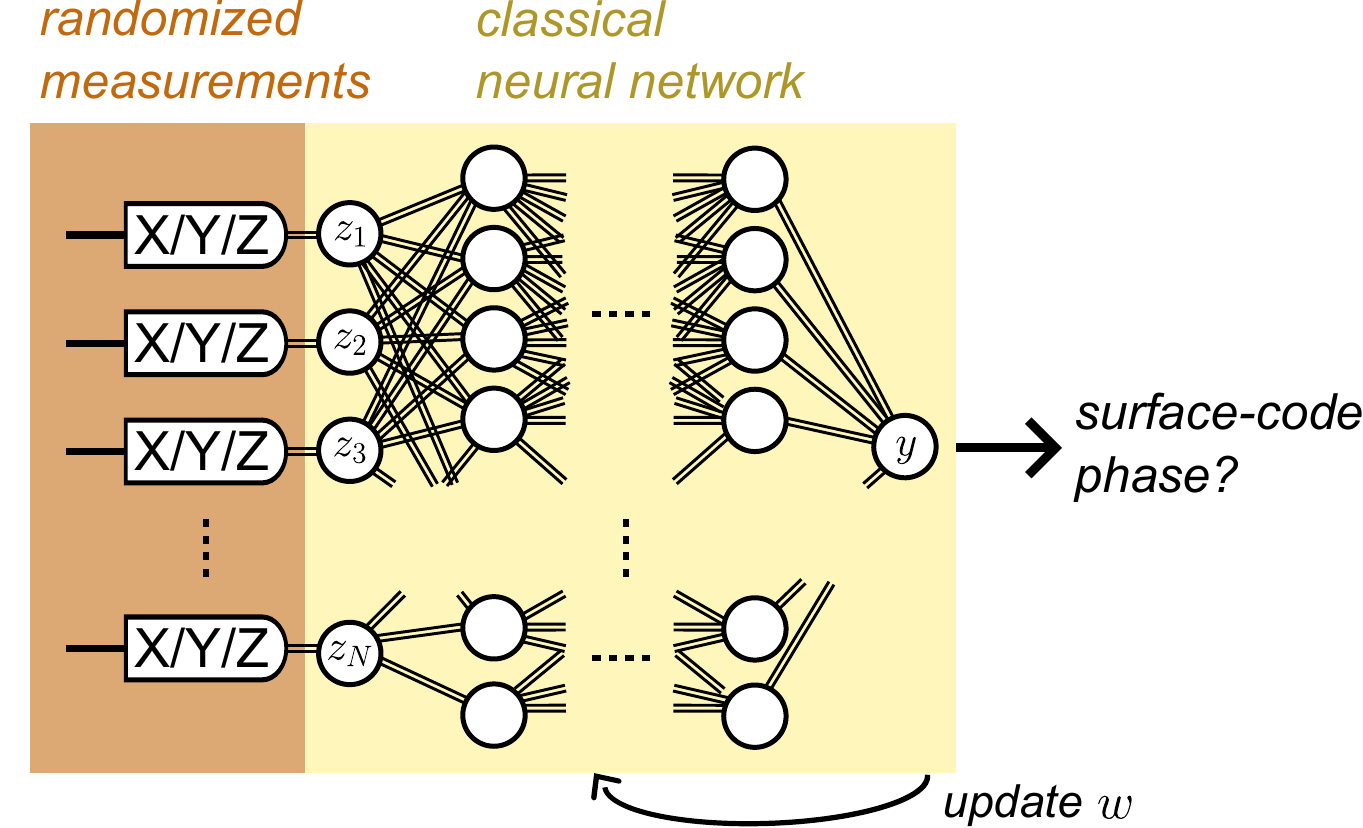}
    \caption{Classical neural network trained on randomized Pauli measurements. Each qubit is measured in a randomly chosen Pauli basis $\sigma$ and the measurement outcome $z=z_1z_2\ldots z_N$ is mapped onto the output $y(z)$ by the classical neural network with weights and biases $w$.}
    \label{Fig:AppC}
\end{figure}

As a benchmark for the hybrid neural network, we consider a classical neural network trained on direct randomized measurements \cite{huang2022}, avoiding any trainable quantum circuit, as shown in Fig.~\ref{Fig:AppC}. 

\subsection{Randomized Pauli-6 POVM measurements}

We employ the Pauli-6 POVM, which is defined for a single qubit $i$ by the elements
$\Pi^{(i)}_{z_i} \in \{\Pi^{(i)}_0 = \frac{1}{3}| + \rangle \langle + |,\Pi^{(i)}_1 =\frac{1}{3}| - \rangle \langle - |, \Pi^{(i)}_2 =\frac{1}{3}| +i \rangle \langle +i |, \Pi^{(i)}_3 =\frac{1}{3}| -i \rangle \langle -i |, \Pi^{(i)}_4 =\frac{1}{3}| 0 \rangle \langle 0 |,\Pi^{(i)}_5 = \frac{1}{3}| 1 \rangle \langle 1 | \}$, where $z_i =0, 1, 2, 3, 4, 5$. For $N$ qubits, the POVM elements are given by the tensor product of the single-qubit POVM elements $\Pi_z = \bigotimes_{i=1}^N \Pi_{z_i}^{(i)}$, where $z=z_1z_2\ldots z_N$. The number of these elements is $6^N$, which is approximately $10^7, 3\times10^{12}, 3\times10^{19}$ for $N=9,16,25$, respectively. 

This POVM is informationally complete and can be implemented by randomized Pauli measurements \cite{carrasquilla2019}. Specifically, we measure in the eigenbases of Pauli strings $\sigma = \bigotimes_{i=1}^N\sigma_i$, where $\sigma_i \in X_i, Y_i, Z_i$. For each training state $\ket{\psi_L^{(m)}}$, we generate a dataset $\mathcal{D}_L^{(m)}$ of outcomes $z$ by repeatedly performing these measurements. In each measurement shot, a Pauli basis $\sigma$ is chosen uniformly at random and the outcome $z$ labels the corresponding eigenstate associated with the POVM element $\Pi_z$.

For this Pauli-6 POVM measurement, we estimate the Jensen-Shannon divergence  $D_{JS}[P_0,P_1]$ by averaging over a set $\mathcal{B}$ of $10^6$ randomly chosen Pauli bases $\sigma$, 
\begin{align}
    D_{JS}[P_0,P_1]&=\frac{1}{3^N}\sum_{\sigma\in\{X,Y,Z\}^{\otimes N}}D_{JS}[P_0^{(\sigma)},P_1^{(\sigma)}]\label{eq:jsd_pauli}\\
    &\approx \frac{1}{|\mathcal{B}|} \sum_{\sigma\in \mathcal{B}}D_{JS}[P_0^{(\sigma)},P_1^{(\sigma)}],
\end{align}
where $P_L(z)$ and $P^{(\sigma)}_L(x)$ are the distributions of the Pauli-6 POVM measurements and the measurements in basis $\sigma$, respectively, defined according to Eq.~\eqref{eq:probclass} with $P_L^{(m)}(z)= \bra{\psi_L^{(m)}} \Pi_z \ket{\psi_L^{(m)}}$. Equality \eqref{eq:jsd_pauli} holds since $P_L(z)=\frac{1}{3^N}P_L^{(\sigma)}(z \mod 2)$, where $\sigma_i=X_i,Y_i,Z_i$ for $\lfloor z_i/2 \rfloor=0,1,2$, respectively. For the training data comprising the surface code in a magnetic field and the 1-design, we estimate $D_{JS}[P_0,P_1]=0.34$ for $N=25$. Analogously, we determine the Bayes error rate $p_{\rm opt}= 16.8\%$. 

\subsection{Classical benchmark network training}
    
The classical feedforward neural network consists of an input layer with $N$ nodes, followed by $l$ fully-connected hidden layers with $n$ nodes and rectified linear unit activation functions, and a single-node output layer with a sigmoid activation function.

Similarly to the hybrid neural network, the classical benchmark network learns linear functions of the input quantum state $\ket{\psi_L^{(m)}}$. In particular, the mean output 
\begin{equation}
\bar{y} = \frac{1}{|\mathcal{D}^{(m)}_L|} \sum_{z\in\mathcal{D}^{(m)}_L} y(z)
\end{equation}
corresponds to the expectation value $\lim_{|\mathcal{D}^{(m)}_L|\rightarrow\infty}\bar{y}=\textrm{Tr}[O\ket{\psi_L^{(m)}}\bra{\psi_L^{(m)}}]$ of the observable $O=\sum_{z\in\{0,1,2,3,4,5\}^N}y(z)\Pi_z$.

The training states and labels are identical to those used for the hybrid neural network, see Appendix~\ref{sec:train}. 
The datasets $\mathcal{D}^{(m)}_L$ are split into training datasets $\mathcal{D}^{(m)}_{L,\textrm{tr}}$ and validation datasets $\mathcal{D}^{(m)}_{L,\textrm{val}}$ with the ratio $|\mathcal{D}^{(m)}_{L,\textrm{tr}}|/|\mathcal{D}^{(m)}_{L,\textrm{val}}|=4$.
The measurement outcomes $z_i$ are standardized as $z_i \mapsto (z_i - \bar{z}_i)/s_i$, where $\bar{z}_i$ and $s_i$ denote the mean and standard deviation of the training datasets $\mathcal{D}^{(m)}_{L,\textrm{tr}}$ for all $L$ and $m$. 

The network parameters $w$ are optimized by minimizing the binary cross-entropy cost \eqref{eq:bxe} over all datasets $\mathcal{D}_L^{(m)}$. In contrast to the hybrid neural network, no measurements are performed during training. The classical neural network is implemented using the Keras API within TensorFlow, and trained via backpropagation with stochastic gradient descent using the Adam optimizer, similarly to the feedforward neural network in our hybrid approach. To mitigate overfitting, we also employ early stopping.

    \begin{figure}[t]
        \centering
        \includegraphics[width=1\linewidth]{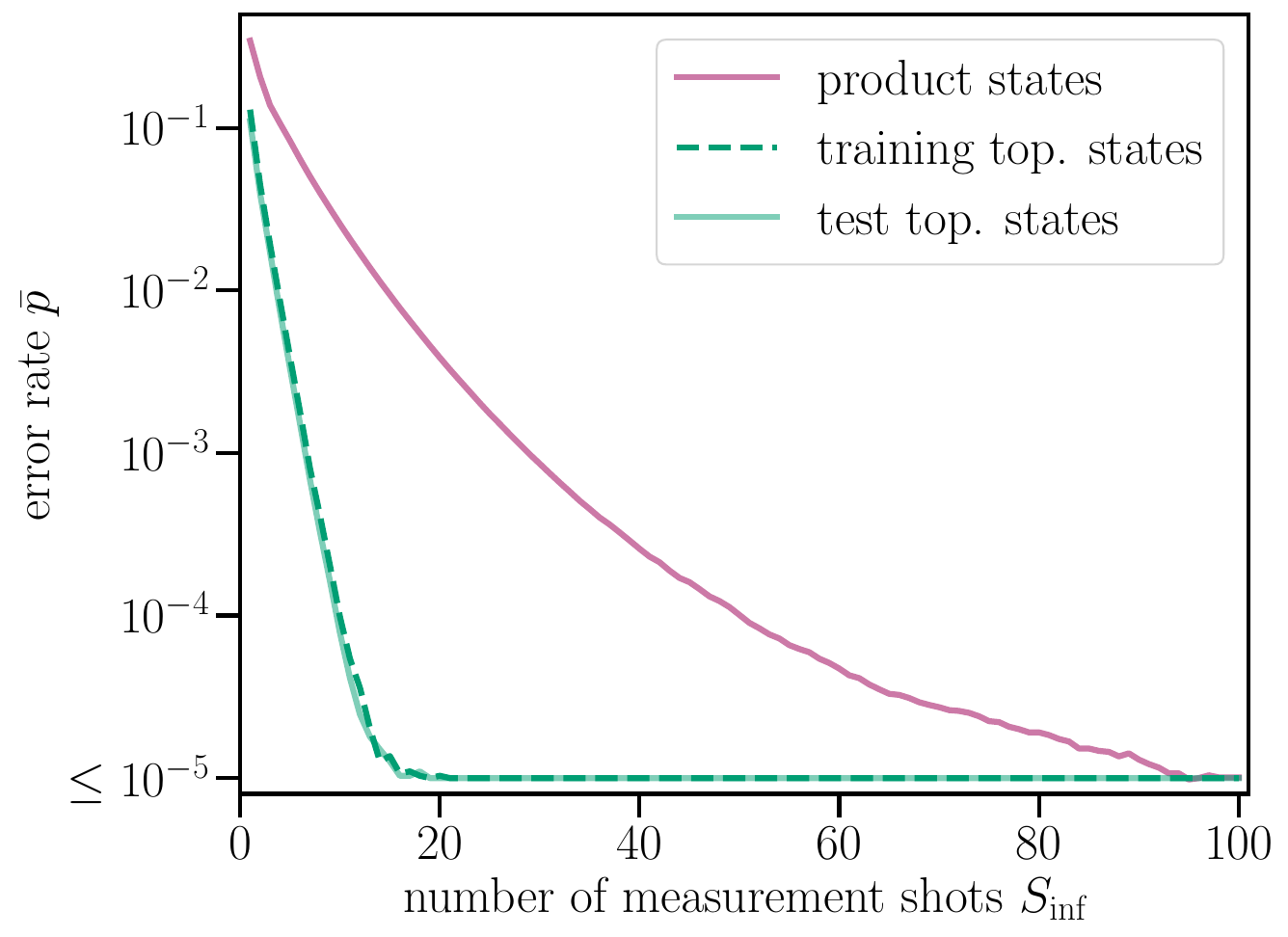}
        \caption{Multi-shot inference for recognizing the topological phase of the surface code in a magnetic field using the classical benchmark network. The false negative error rate $\bar{p}_{\rm fn}$ for training topological states (dashed green line) and test topological states (solid green line), as well as the false positive error rate $\bar{p}_{\rm fp}$ for $10^7$ random product states (purple line) as a function of the measurement shot number $S_{\rm inf}$ in inference.}        
        \label{fig:POVM_error_rates}
    \end{figure}

\begin{figure}[b]
    \centering
    \includegraphics[width=\linewidth]{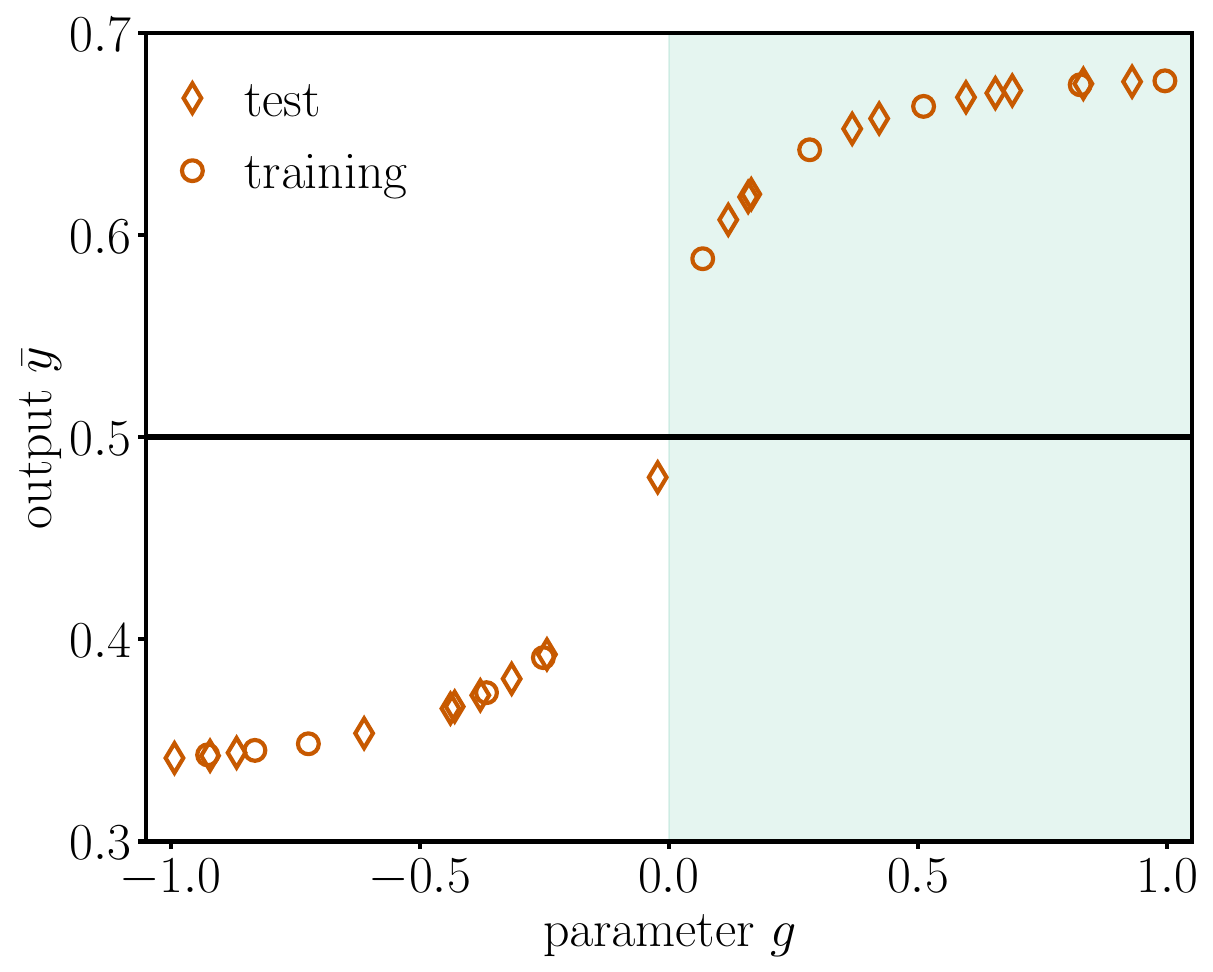}
    \caption{Distinguishing the surface-code phase from the symmetry-enriched topological phase by the classical benchmark network. Mean output $\bar{y}$ as a function of the parameter $g$ for the training states (circles) and test states (diamonds). The shaded region corresponds to the surface-code phase and the decision threshold $t=0.5$ is shown as the black line.}
    \label{fig:tns_povm_output}
\end{figure}

We first tune the hyperparameters of the network for the surface code in a magnetic field. For $N=9$, we perform one-dimensional parameter sweeps over the number of hidden layers  $l= 1,2,3,4,5$, the number of nodes per hidden layer $n = 64, 128, 256, 512, 1024$, and the learning rate $\lambda = 0.001,0.01$. We use 1000 batches and $N_{\rm epochs}=1000$ epochs. Each hyperparameter is individually varied while the other hyperparameters are kept fixed at their baseline values. The hyperparameters $l=3$, $n=256$ and $\lambda=0.001$ yield the lowest validation cost. Similarly, we tune hyperparameters $l=3$, $n=256$, $\lambda=0.001$, and the batch size $N_{\rm batch} = 10^4$ for $N=16$ by sweeping over $l= 1,2,3,4,5$, $n = 128, 256, 512, 1024$, $\lambda = 0.0005, 0.001,0.002$ and $N_{\rm batch} = 10^3, 10^4, 10^5$, where the training is terminated either by early stopping or upon reaching a wall-clock time of approximately 24 hours. For $N=25$, we obtain $l=5$, $n=256$, $\lambda=0.001$, and $N_{\rm batch} = 10^4$ by sweeping 
over $l = 2,3,4,5,6,7,8 $,  $n = 128, 256, 512$, $\lambda = 5\times10^{-5}, 0.0001, 0.0005, 0.001, 0.002 $ and $N_{\rm batch} = 10^3, 10^4, 10^5, 10^6$.

For the tensor-network states spanning the phase transition between the surface-code phase and the symmetry-enriched topological phase, one-hot encoding of the measurement outcomes $z$ facilitates the training of the network. Specifically, each outcome $z_i$ is encoded as a vector of length 6 with a single nonzero entry equal to 1, whose position uniquely identifies $z_i$. Concatenating these encodings for all qubits yields a vector of length $6N$, which is standardized and used as input to the benchmark network. Accordingly, the number of nodes in the input layer is increased to $6N$. To tune the hyperparameters, we sweep them over the following values: $l= 1,2,3,4,5,6,7,8$, $n = 128, 256, 512$, $\lambda = 0.00025, 0.0005, 0.001, 0.002$ and $N_{\rm batch} = 10^3, 10^4, 10^5, 10^6$. The lowest validation cost is achieved for $l=5$, $n=256$, $N_{\rm epochs}=200$, $N_{\rm batch} = 10^4$.

\subsection{Topological phase recognition}
\begin{figure}[t]
    \centering
    \includegraphics[width=\linewidth]{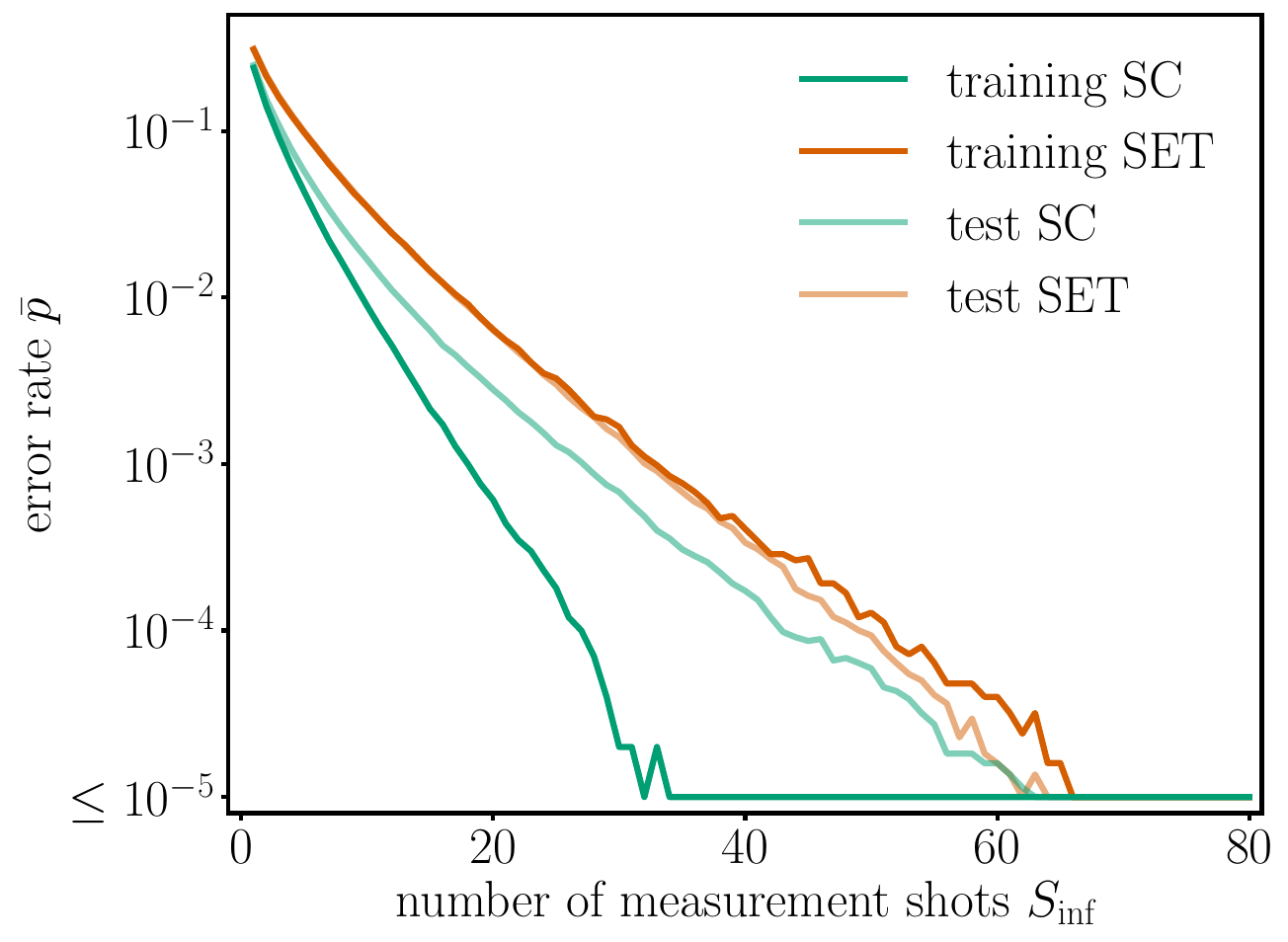}
    \caption{Multi-shot inference for recognizing the surface-code (SC) phase and the symmetry-enriched topological (SET) phase with the classical benchmark network. The false negative error rate $\bar{p}_{\rm fn}$ for the training SC states (dark green line) and test SC states (light green line), as well as the false positive error rate $\bar{p}_{\rm fp}$ for the training SET states (dark red) and the test SET states (light red) as a function of the measurement shot number $S_{\rm inf}$.
    }
    \label{fig:tns_povm_inference}
\end{figure}
First, we investigate the phase classification performance of the benchmark network trained on $S_{\rm tot} = 10^8$ randomized measurements of the surface code in a magnetic field and the 1-design for $N=25$. In Fig.~\ref{fig:POVM_error_rates}, we plot the multi-shot false negative error rate $\bar{p}_{\rm fn}$ as a function of the inference shot number $S_\mathrm{inf}$ for the training topological states (dashed green line) and test topological states (solid green line). The false positive error rate $\bar{p}_{\rm fp}$ evaluated on $10^7$ random product states falls below $10^{-4}$ for $S_{\mathrm{inf}} = 50$ inference shots, see the purple line in  Fig.~\ref{fig:POVM_error_rates}.

We now study the phase classification performance of the benchmark network trained on $S_{\rm tot}=10^8$ randomized measurements of the tensor-network states spanning the transition between the surface-code phase and the symmetry-enriched topological phase. In Fig.~\ref{fig:tns_povm_output}, we plot the mean output $\bar{y}$ of the trained benchmark network as a function of the parameter $g$. We can see that the network correctly classifies both the training states (circles) and test states (diamonds). However, this classification requires a larger number of measurement shots $S_{\rm inf}$ than for the hybrid neural network discussed in Sec.~\ref{sec:set}. In particular, we can see in Fig.~\ref{fig:tns_povm_inference} that $S_{\rm inf}=52$ is required to achieve near-perfect classification with both the false positive error rate $\bar{p}_{\rm fp}<0.01\%$ (red lines) and the false negative error rate $\bar{p}_{\rm fn}<0.01\%$ (green lines). In this plot, we exclude states for $|g|<0.15$ lying close to the phase transition. In contrast, the hybrid neural network reaches comparable error rates with only $S_{\rm inf} = 13$.

\bibliography{HNN_paper}

\end{document}